\documentclass[prb,showpacs,twocolumn,superscriptaddress,floatfix]{revtex4-1}

\usepackage{epsfig}
\usepackage{amsmath}
\usepackage{amssymb}
\usepackage{amsfonts}
\usepackage{mathptmx}
\usepackage{dcolumn}
\usepackage{eucal}
\usepackage{bm}
\usepackage{color}
\usepackage[colorlinks,linkcolor=blue,citecolor=blue]{hyperref}
\usepackage{epstopdf}
\usepackage{bbold}
\usepackage{url}

\usepackage{dsfont}

\def\eps{\varepsilon}

\def\bk{{\bf k}}

\def\up{\uparrow}
\def\dw{\downarrow}

\def\be{\begin{equation}}
\def\ee{\end{equation}}
\def\bea{\begin{eqnarray}}
\def\eea{\end{eqnarray}}
\def\ba{\begin{array}{l l}}
\def\ea{\end{array}}

\begin{document}

\title{Electrically Tunable Superconductivity Through Surface Orbital Polarization}

\author{Maria Teresa Mercaldo}
\affiliation{Dipartimento di Fisica ``E. R. Caianiello", Universit\`a di Salerno, IT-84084 Fisciano (SA), Italy}

\author{Paolo Solinas}
\affiliation{SPIN-CNR, Via Dodecaneso 33, 16146 Genova, Italy}
\affiliation{Dipartimento di Fisica, Universit\'a di Genova and INFN Sezione di Genova, Via Dodecaneso 33, 16146 Genova, Italy}

\author{Francesco Giazotto}
\affiliation{NEST, Istituto Nanoscienze-CNR and Scuola Normale Superiore, Piazza San Silvestro 12, I-56127 Pisa, Italy
}

\author{Mario Cuoco}
\affiliation{SPIN-CNR, IT-84084 Fisciano (SA), Italy}
\affiliation{Dipartimento di Fisica ``E. R. Caianiello", Universit\`a di Salerno, IT-84084 Fisciano (SA), Italy}

\begin{abstract}
We investigate the physical mechanisms for achieving an electrical control of conventional spin-singlet superconductivity in thin films by focusing on the role of surface orbital polarization. Assuming a multi-orbital description of the metallic state, due to screening effects the electric field acts by modifying the strength of the surface potential and, in turn, yields non-trivial orbital-Rashba couplings. The resulting orbital polarization at the surface and in its close proximity is shown to have a dramatic impact on superconductivity. We demonstrate that, by varying the strength of the electric field, the superconducting phase can be either suppressed, i.e. turned into normal metal, or undergo a $0-\pi$ transition with the $\pi$ phase being marked by non-trivial sign change of the superconducting order parameter between different bands. These findings unveil a rich scenario to design heterostructures with superconducting orbitronics effects.
\end{abstract}
\maketitle

\section{Introduction}
Because of the screening effect, a static electric field (EF) cannot penetrate inside a metal deeper than a few Thomas-Fermi lengths ($0.1-1~$nm) \cite{Ashcroft, Lang1970, UmmarinoPhysRevB2017}.
As a consequence, the behaviors and features of a metal, e.g., its transport properties, are practically unaffected by the application of static EFs.

Analogously, when dealing with the interaction of a static EF with a superconductor (SC) \cite{Shapiro1984, LipavskyPhysRev2002, Koyama2001, MachidaPhysRevLett2003},
for standard metallic SCs, that are well described by the Bardeen-Cooper-Schrieffer theory \cite{deGennes,tinkham2012introduction}, the penetration length of an EF is roughly unchanged with respect to the normal metal phase \cite{virtanen2019superconducting}.
In this context, recent experiments have shown that a strong static EF can dramatically affect the properties of superconducting wires and planes 
\cite{DeSimoniNatNano2018, PaolucciNanoLett2018, PaolucciPhysRevAppl2019, DeSimoni2019mesoscopic, Paolucci2019connecting} suppressing the supercurrent, and inducing a superconductor-to-normal metal transition.
This superconducting field effect (SFE) is quite ubiquitous since it has been observed in different materials \cite{DeSimoniNatNano2018,desimoni1}, in Dayem bridges \cite{PaolucciNanoLett2018, PaolucciPhysRevAppl2019}, in superconductor-normal metal-superconductor mesoscopic junctions \cite{DeSimoni2019mesoscopic}, and in 
superconducting quantum interference devices \cite{Paolucci2019connecting}.
Hence, these experimental evidences suggest that the SFE is a genuine phenomenon which cannot be explained in terms of well-known effects such as charge accumulation or depletion \cite{PaolucciPhysRevAppl2019, Paolucci2019connecting}.

A basic remark is that the Cooper pairs are correlated over distances ($\xi_0$) much longer than the EF screening length and thus a perturbation occurring at the edge of the superconductor may affect the system within a distance comparable to $\xi_0$.
This expectation seems to be confirmed by the fact that the SFE is observable only on structures with characteristic dimensions of a few coherence lengths, and then vanishes exponentially \cite{DeSimoniNatNano2018}.
Besides this, our understanding of the physics at the origin of the SFE is somewhat limited \cite{DeSimoniNatNano2018, PaolucciPhysRevAppl2019, Paolucci2019connecting}, and a fully microscopic theory is still missing.

Motivated by the above experimental results \cite{DeSimoniNatNano2018, PaolucciNanoLett2018, PaolucciPhysRevAppl2019, DeSimoni2019mesoscopic, Paolucci2019connecting}, in this paper we propose a theoretical model which is able to grasp some of the observed features typical of the SFE and to provide a microscopic physical scenario to account for the modification of the superconducting order parameter (OP) due to the applied EF at the surface.
Our key idea is to consider the effects of the EF as a source of inversion symmetry breaking at the surfaces of the superconductor and to focus on the consequences of the induced orbital polarization on the electron pairing. 
It has been recently recognized that an orbital analogue of the spin Rashba effect \cite{Rashba1960} can be achieved on the surfaces \cite{Park2013,Kim2014,Petersen2000} even in the absence of atomic spin-orbit coupling \cite{Go2017}. The orbital Rashba (OR) interaction allows for mixing of orbitals on neighboring atoms that would not overlap in an inversion symmetric configuration. 
Such coupling leads to non-vanishing orbital polarization that form chiral patterns in the momentum space. Remarkably, the OR coupling is quite ubiquituous in metals and semiconductors since it occurs either in pure $p$- and $d$-orbitals \cite{Park2013,Kim2014,Petersen2000} or $sp$- or $pd$-hybridized systems \cite{Go2017}.
Evidences of anomalous electronic splitting and of the role played by the orbital degrees of freedom have been found on a large variety of surfaces~\cite{el-kareh14}, Bi/Ag(111)~\cite{schirone15}, etc.\, as well as in oxide interfaces~\cite{King2014,Nakamura2012,Fukaya2019}.

\begin{figure}[t!]
\centering
\includegraphics[width=0.4\textwidth]{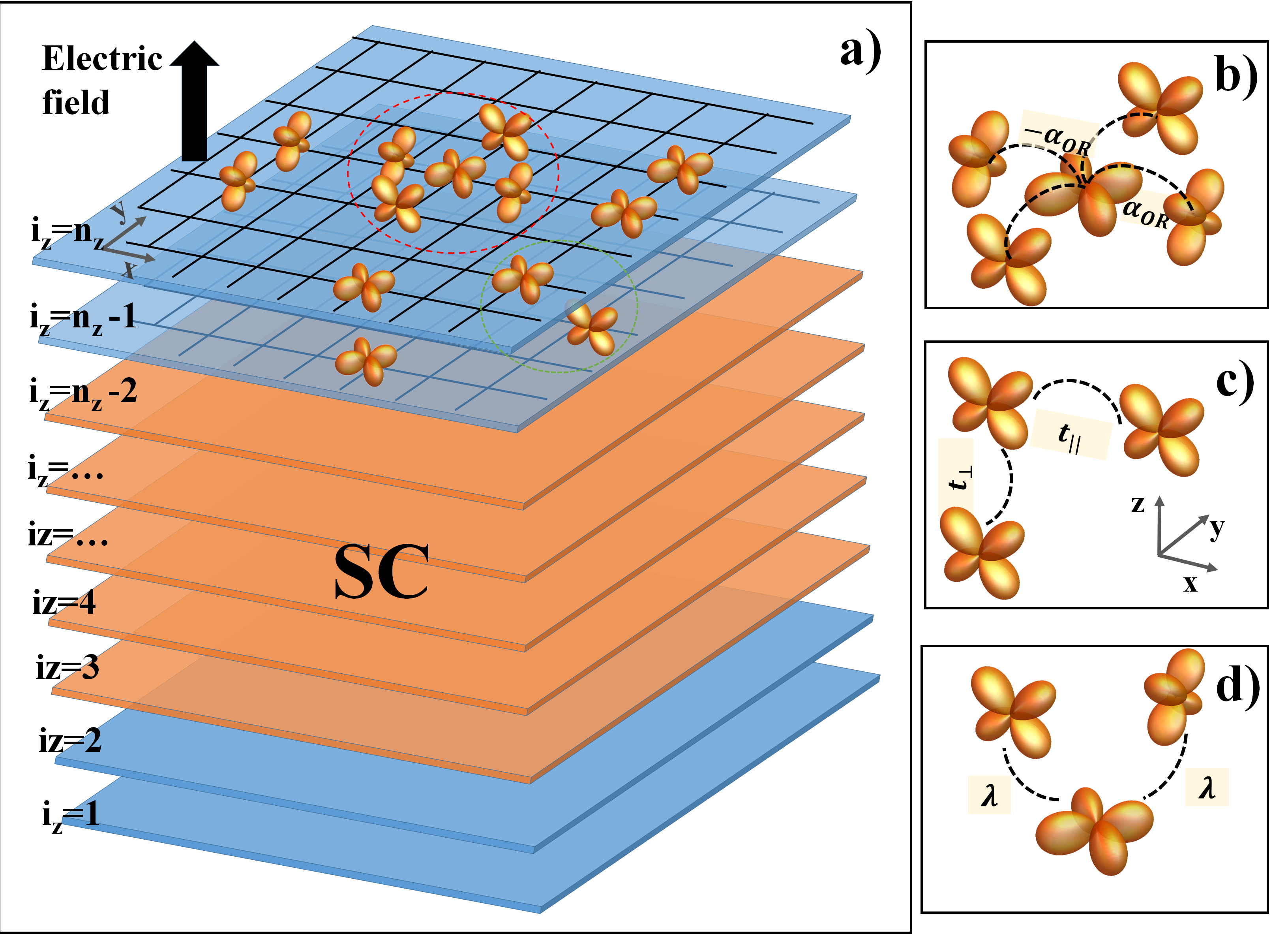}
\protect\caption{(a) Schematic view of multilayered spin-singlet superconductor (SC) with $n_z$ layers labelled by the index $i_z$. The electric field penetrates only at the surface layers (blue) by inducing processes with intra and inter-layer orbital mixing. In the remaining layers (orange), the electric field is absent. (b) Sketch of the surface electronic hybridization due to orbital Rashba coupling, $\alpha_{OR}$, between $xy$ and 
$(xz,yz)$ orbitals along the symmetry allowed directions. The standard nearest neighbor hopping between $d$-orbitals is mainly relevant for homologue orbitals along $x,y,z$ axes, e.g. in (c) $xz$ orbitals hybridize along $x$ and $z$ directions, with $t_{||}$ and $t_{\perp}$, respectively. Panel (d) depicts the effect of $\lambda$ with orbital mixing involving $xy$ and $xz,yz$ states. The term $\lambda$ is active only between the first two surface layers (blue). The orbital Rashba coupling is considered to be non vanishing only at the surface layers.}
\label{Fig:1}
\end{figure}

Here, we consider how the induced orbital polarization at the surface is able to significantly modify the amplitude and phase of conventional spin-singlet superconducting OP in thin films. Through a multi-orbital description we show that the EF can suppress the superconducting state at the surface by inducing a substantial orbital polarization close to the Fermi level.
Then, the occurrence of orbitally polarized surface states can guide a complete breakdown of the superconducting state in the whole system or an unconventional $0$-$\pi$ transition with a non-trivial sign change of the superconducting OP between different bands. Although this phase resembles the unconventional $s_{\pm}$ pairing proposed in iron based superconductors~\cite{spm1,spm2,spm2}, our analysis has a completely different root since it demonstrates that the EF can stabilize a $\pi$-phase in conventional $s-$wave superconductors. The resulting phase transitions manifest themselves as a consequence of the interplay of two fundamental electronic processes which we microscopically demonstrate to arise from the surface electrostatic potential (Appendix A): i) intra-layer  $\alpha_{OR}$, ii) inter-layer $\lambda$ OR interactions, respectively (Fig. \ref{Fig:1}). 
Both $\alpha_{OR}$ and $\lambda$ are proportional to the strength of electric field, $E_s$, with $\lambda$ being generally smaller than $\alpha_{OR}$ and activated by in-plane atomic distortions or strain effects.
Our study thus uncovers fundamental mechanisms for an electrical control of conventional superconductors based on the modification of the orbital polarization at the surface.  

The paper is organized as follows. In Sect. II we provide the basic elements of the modelling and of the methodology. Sect. III is devoted to the main results including the phase diagram and the role of pairing interaction, inter-orbital mixing and inter-layer hopping. In Sect. IV we have the concluding remarks and the discussion.
Finally, in the Appendix we provide the derivation of the orbital Rashba couplings due to the surface
electrostatic potential and the impact of the orbital Rashba coupling on superconductivity for a monolayer. Furthermore, we also present the character of the phase transitions by inspecting the
free energy profile in the various regimes, and the behavior of the layer dependent orbital polarization.  

\begin{figure*}[t!]
\centering
\includegraphics[width=0.9\textwidth]{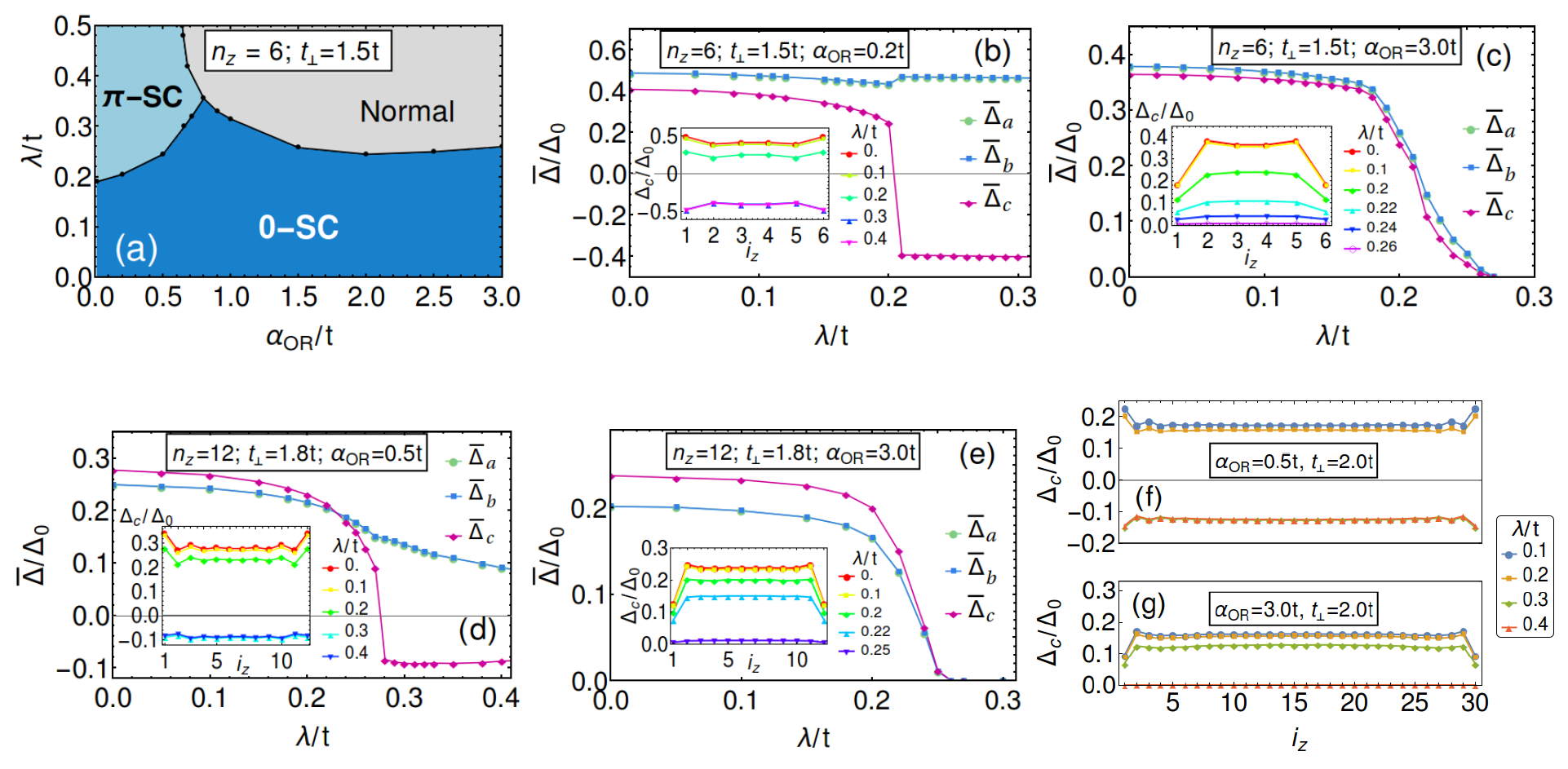}
\protect\caption{(a) Phase diagram in the ($\alpha_{OR},\lambda$)-plane with conventional superconducting (0-SC), unconventional ($\pi$-SC), and normal state. The parameters are:  $n_z=6, \mu=-0.4t, t_\perp=1.5t, \eta=0.1$. 
(b)-(c) Behavior of the order parameter $\bar{\Delta}_\alpha \; (\alpha=a,b,c)$ in the central layer at $i_z=n_z/2$, as function of $\lambda$ in the regimes of weak (panel (b)) and strong (panel (c)) orbital Rashba interaction, namely $\alpha_{OR}=0.2t$ and $\alpha_{OR}=3.0t$, respectively.  In (b) we observe a sharp transition to the $\pi$-SC, with a sign change in $\Delta_c$ and hence a relative $\pi$-phase between $c$ and $a,b$ OPs. In (c) we demonstrate that all the OPs go to zero. 
Insets: $\Delta_c$ along the $\hat{z}$ direction is shown for different values of $\lambda$.
In (d) and (e) we show the analogous transitions of (b) and (c), but for $n_z=12$. In (f) and (g) we present the profile of $\Delta_c$ for $n_z=30$, for weak and strong $\alpha_{OR}$, respectively. In (f) we show the sign change (for $\lambda=0.3t$ and $0.4t$), while in (g) the OP is suppressed by increasing $\lambda$. In (b)-(g) $\Delta_0$ is the superconducting OP for a monolayer without the OR and sets its scale (Appendix B). 
}
\label{Fig:2}
\end{figure*}

\section{Model and methodology} We assume a conventional $s-$wave spin-singlet pairing for a geometry with $n_z$ layers (Fig. \ref{Fig:1}). The electronic description is based on $d$-orbitals, i.e. ($yz,xz,xy$). Since $E_s$ on the surface is parallel to $\hat{z}$, it can be described by a potential $V_{s}=-E_{s} z$.  Following the approach already applied to derive the surface orbital Rashba coupling \cite{Park2011,Park2012,Kim2013}, the matrix elements of $V_s$ in the Bloch basis yield an intra- ($\alpha_{OR}\sim E_s)$ and inter-layer ($\lambda \sim E_s$) inversion asymmetric interactions, whose ratio depends only on the inter-atomic distances and distortions at the surface.
For convenience we indicate as $(a,b,c)$ the ($yz,xz,xy$) $d-$orbitals. Then, we introduce the creation $d^\dagger_{\alpha,\sigma}(\bk,i_z)$ and annihilation $d_{\alpha,\sigma}(\bk,i_z)$ operators with momentum $\bk$, spin ($\sigma=[\up,\dw]$), orbital ($\alpha=(a,b,c$)), and layer $i_z$, to construct a spinorial basis
$\Psi^\dagger(\bk,i_z)=(\Psi_{\up}^\dagger(\bk,i_z), \Psi_{\dw}(-\bk,i_z))$ with
$\Psi_{\sigma}^\dagger(\bk,i_z)=(d^\dagger_{a,\sigma}(\bk,i_z),d^\dagger_{b,\sigma}(\bk,i_z),d^\dagger_{c,\sigma}(\bk,i_z))$. In this representation, the Hamiltonian can be expressed in a compact way as:
\begin{eqnarray}
\mathcal{H}= \frac{1}{N} \sum_{\bk,i_z,j_z} \Psi^{\dagger}(\bk,i_z) \hat{H}(\bk) \Psi(\bk,j_z) \,,
\end{eqnarray}
\noindent with 
\begin{eqnarray}
&& \hat{H}(\bk)=\sum_{\alpha=\{a,b,c\}} [\tau_z \eps_\alpha(\bk) + \Delta_\alpha(i_z) \tau_x ]\otimes(\hat{L}^2 - 2 \hat{L}^2_\alpha)] \delta_{i_z,j_z} + \nonumber \\
&&+ \alpha_{OR} \tau_z \otimes (\sin k_y \hat{L}_x - \sin k_x \hat{L}_y) [\delta_{i_z,j_z}(\delta_{i_z,1} + \delta_{i_z,n_z})] + \nonumber \\
&& + t_{\perp,\alpha} \tau_z \otimes (\hat{L}^2 - 2 \hat{L}^2_\alpha) \delta_{i_z,j_z\pm 1}+\nonumber \\
&& +  \lambda \left[(\hat{L}_x+\hat{L}_y) (\delta_{i_z,1}\delta_{j_z,2}+\delta_{i_z,n_z}\delta_{j_z,n_z-1}) +\text{h.c.} \right] \,,
\end{eqnarray}
\noindent where the orbital angular momentum operators $\hat{L}$ have components 
\begin{eqnarray*}
\hat{L}_x=\begin{bmatrix}
0 & 0 & 0 \\ 
0 & 0 & i \\ 
0 & -i & 0%
\end{bmatrix}, 
\hat{L}_y=\begin{bmatrix}
0 & 0 & -i \\ 
0 & 0 & 0 \\ 
i & 0 & 0%
\end{bmatrix}, \hat{L}_z=\begin{bmatrix}
0 & -i & 0 \\ 
i & 0 & 0 \\ 
0 & 0 & 0%
\end{bmatrix}
\end{eqnarray*}
\noindent within the ($yz,xz,xy$) subspace, $\tau_i$ ($i=x,y,z$) are the Pauli matrices for the electron-hole sector, and $\delta_{i,j}$ the Kronecker delta function. 
The kinetic energy for the in-plane electron itinerancy is due to the symmetry allowed \cite{Slater1954} nearest neighbor hopping, thus, one has that $\eps_a(\bk)=-2 t_{||} [\eta \cos(k_x)+\cos(k_y)]$, $\eps_b(\bk)=-2 t_{||} [\cos(k_x)+ \eta \cos(k_y)]$, and $\eps_c(\bk)=-2 t_{||} [\cos(k_x)+ \cos(k_y)]$, with $\eta$ being a term that takes into account deviations from the ideal cubic symmetry. The role of inter-orbital hopping that are activated by distortions has been explicitly evaluated. 
We assume that the layer dependent spin-singlet OP is non-vanishing only for electrons belonging to the same band and it is expressed as $\Delta_\alpha(i_z)=\frac{1}{N} \sum_{\bk} g\,\langle d_{\alpha,\uparrow}(\bk,i_z) d_{\alpha,\downarrow}(-\bk,i_z)  \rangle$ with $\langle ...\rangle$ being the expectation value on the ground state. 
Here, $N=n_x \times n_y$ sets the dimension of the layer in terms of the linear lengths $n_x$ and $n_y$, while we assume translation invariance in the $xy$-plane and $n_z$ layers along the $z-$axis (Fig. \ref{Fig:1}). We point out that $g$ is not modified by the electric field. This is physically consistent with the fact that due to screening effects the EF cannot induce an inversion asymmetric potential inside the thin film beyond the Thomas-Fermi length.
The analysis is performed by determining the superconducting OPs corresponding to the minimum of the free energy employing a self-consistent iterative procedure until the desired accuracy is achieved.
The planar hopping is the energy unit, $t_{||}=t$, while the interlayer one is orbital independent, i.e. $t_{\perp,\alpha}=t_{\perp}$. Within the same scheme of computation we also consider the role of amplitude's variation of the intra-orbital pairing interaction and of the inter-orbital superconducting interaction, $g_{od}$, with the corresponding OPs $\Delta_{\alpha\beta}$ with $\alpha \neq\beta$. Here, the inter-orbital OPs are expressed as $\Delta_{\alpha\beta}(i_z)=\frac{1}{N} \sum_{\bk} g_{{od}} \,\langle d_{\alpha,\uparrow}(\bk,i_z) d_{\beta,\downarrow}(-\bk,i_z)  \rangle$.

\section{Results} 

In this Section we present the phase diagram as due to the OR couplings and analyze the impact of the pairing interaction, the inter-orbital mixing and the inter-layer hopping.
The effect of the OR couplings is to induce an orbital polarization at the surface and to form chiral orbital textures in the Brillouin zone close to the Fermi level (Appendix D). Moreover, the orbital polarization is generally associated to a configuration with non vanishing angular momentum components and thus it tends to reduce the superconducting OP amplitude (Appendix B) assuming that the pairing interaction preserves inversion symmetry.  
Both interlayer electronic processes, i.e. $\lambda$ and $t_{\perp}$, allow for a transfer of orbital polarization into the inner layers of the superconducting films. Further, due to the symmetry of the orbital processes induced by $\lambda$, there is a drive to develop an orbital dependent phase of the superconducting OP. This aspect can be deduced by evaluating and deducing the behavior of the inter-orbital superconducting OP when $\lambda$ and $\alpha_{OR}$ are the only orbital mixing terms. 

\subsection{Phase diagram}
To get more insight into the role of the electric field it is instructive to start with the phase diagram of the heterostructure for the $n_z=6$ multilayer in the absence of hoppings and pairing terms that mix the orbitals. Considering that a variation of the electric field $E_s$ tunes the interactions $\alpha_{OR}$ and $\lambda$ (Fig. \ref{Fig:2}(a)) we scan the whole amplitude phase space. The outcome is presented for a representative value of the out-of-plane hopping ($t_{\perp}=1.5 t$). 
The conventional superconducting state ($0-$SC), depending on the ratio $\alpha_{OR}/\lambda$, undergoes a transition into two distinct phases: i) an unconventional $\pi-$phase with non-trivial superconducting phase relation between the orbital dependent OPs for a ratio about smaller than one-half, otherwise ii) a normal metal configuration with a vanishing superconducting OP. The nature of the phase transitions can be tracked by following the layer and orbital dependent behavior of $\Delta_\alpha(i_z)$. In the regime of weak $\alpha_{OR}$ the increase of $\lambda$ leads to a complete reconstruction of the superconducting phase. We find that there is a first order phase transition (Appendix C) between two superconducting phases with a reorganization of the relative phase between the orbital dependent OPs. As demonstrated in Fig. \ref{Fig:2}(b), at a critical value of $\lambda$ the superconducting OP for the $c$-band undergoes a first order phase transition with an abrupt sign change of $\Delta_c(i_z)$ in all the layers (see inset Fig. \ref{Fig:2}(b)) while the other two OPs exhibit a discontinuous variation of the amplitude which is sign conserving. The sign change of the OP for one of the band implies an inter-orbital $\pi$-phase between the electron pairs within the $(a,b)$ and $c$ orbitals. Such an orbital reconstruction is an evidence of an unconventional pairing which can directly manifest in an anomalous Josephson coupling with non-standard current-phase relations. The fact that the band $c$ undergoes a sign change of the OP with respect to the $a,b$ bands is a consequence of the structure of the asymmetric inversion couplings at the interface which allow for orbital mixing between $c$ and $(a,b)$ bands. 
The presence of competing phases is also evident if one considers the free energy dependence of the superconducting OP. Indeed, in order to catch the main competing mechanisms, one can assume a uniform spatial profile as a function of the layer index by allowing for an orbital dependent phase reconstruction of the type $\Delta_{\alpha}(i_z)=\exp[i \phi_\alpha] \Delta_0$. Hence, one can directly observe two distinct minima in the free energy, associated with the 0- and $\pi$ phases, whose relative energy difference can be tuned by varying the amplitude of $\lambda$ (Appendix C).

Moving to a larger value of the OR coupling (i.e. $\alpha_{OR}/t \geq 1$) the surface inter-layer coupling $\lambda$ is able to suppress the superconducting state by vanishing the OP amplitude (Fig. \ref{Fig:2}(c)). The value of the critical $\lambda$ setting the 0-SC/normal boundary has a maximum at $\alpha_{OR}/t \sim 1$ and then stays about unchanged by further increasing the OR coupling. Such behavior is accompanied by a qualitative change of the superconducting OP at the surface which starts to get reduced once $\alpha_{OR}$ induces a sufficiently large orbital polarization nearby the Fermi level. The breakdown of the superconductivity in this regime is linked to the character of the Cooper pairs having non-vanishing ${\hat{L}^2}_\alpha$ (i.e. inversion symmetry is preserved), while the EF leads to a large orbital polarization at the surface whose leaking into the inner layers suppresses the pairing amplitude. 
The 0-SC/normal metal phase transition appears to be continuous and it occurs about simultaneously for all the orbitals involved in the pairing close to the Fermi level (Fig. \ref{Fig:2}(c)). It is interesting to notice that a closer inspection of the free energy profile with suitably selected boundary conditions of the OPs at the surfaces and uniform spatial profile in the other layers indicates a smeared type of phase transition from superconductor-to-normal state with weak first order precursors due to the competition between OP configurations with inequivalent amplitude (see Appendix C for details). This implies that the breakdown of the superconducting state, as driven by $\lambda$, is different from that which can be obtained in a standard BCS thermal evolution of the OP. 

After having fully addressed the most favorable superconducting configurations in a thin film with $n_z=6$ layers, we consider whether the orbital asymmetric potential at the surface is able to be also effective in thicker layered films. Such issue is accounted by simulating the cases with $n_z=12$ and $n_z=30$. In Figs. \ref{Fig:2}(d),(e) we demonstrate that for two representative values of $\alpha_{OR}$, corresponding to weak and strong orbital Rashba couplings, the surface interlayer interaction is able to induce the 0-$\pi$ and superconductor-normal metal phase transitions. The phase diagram and the effects are then confirmed and observable either for doubling the system size, $n_z=12$ (Figs. \ref{Fig:2}(d),(e) or for superconducting thin film with $n_z=30$ layers (Figs. \ref{Fig:2}(f),(g)). However, one remark is relevant here concerning the amplitude of the kinetic energy along the $z$-axis. Indeed, the change of the superconducting state is related to the inter-layer hopping amplitude and one needs a slighlty larger $t_{\perp}$ to get critical boundaries occurring in the same range of strengths for $\lambda$ as for thinner SCs (Sect. III E). 


\subsection{Role of the pairing interaction strength}

\begin{figure*}[bt]
\includegraphics[width=0.19\textwidth]{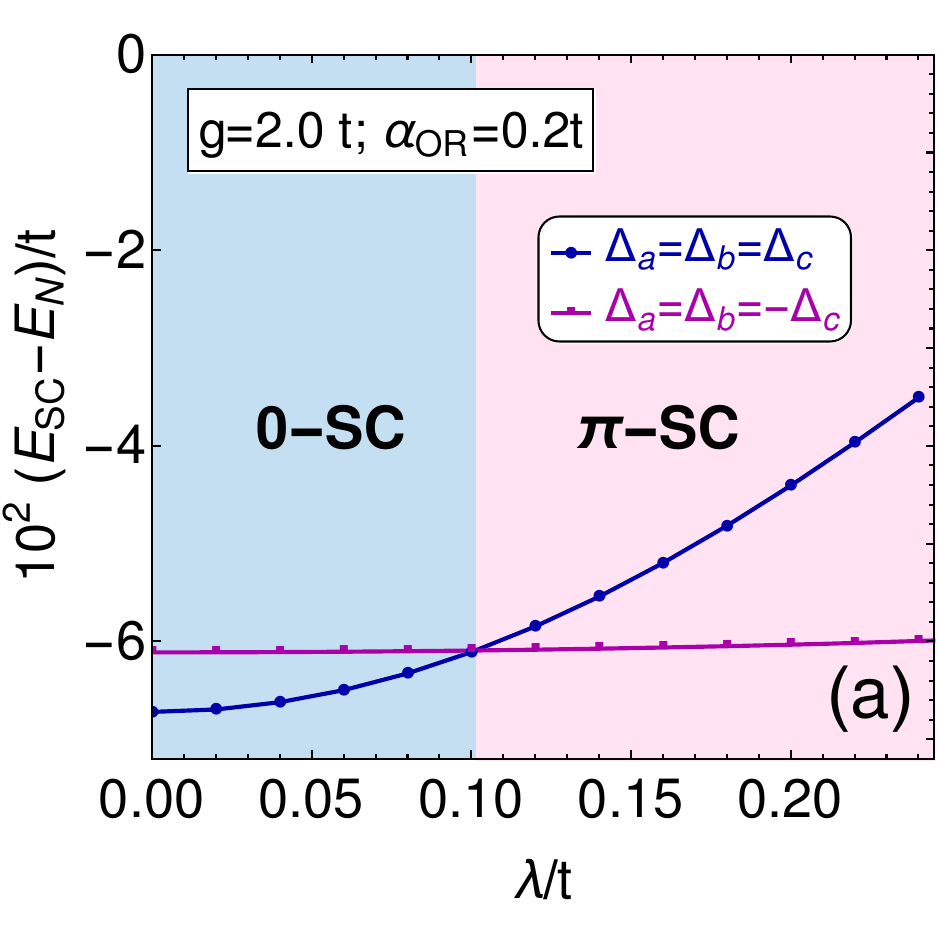}
\includegraphics[width=0.19\textwidth]{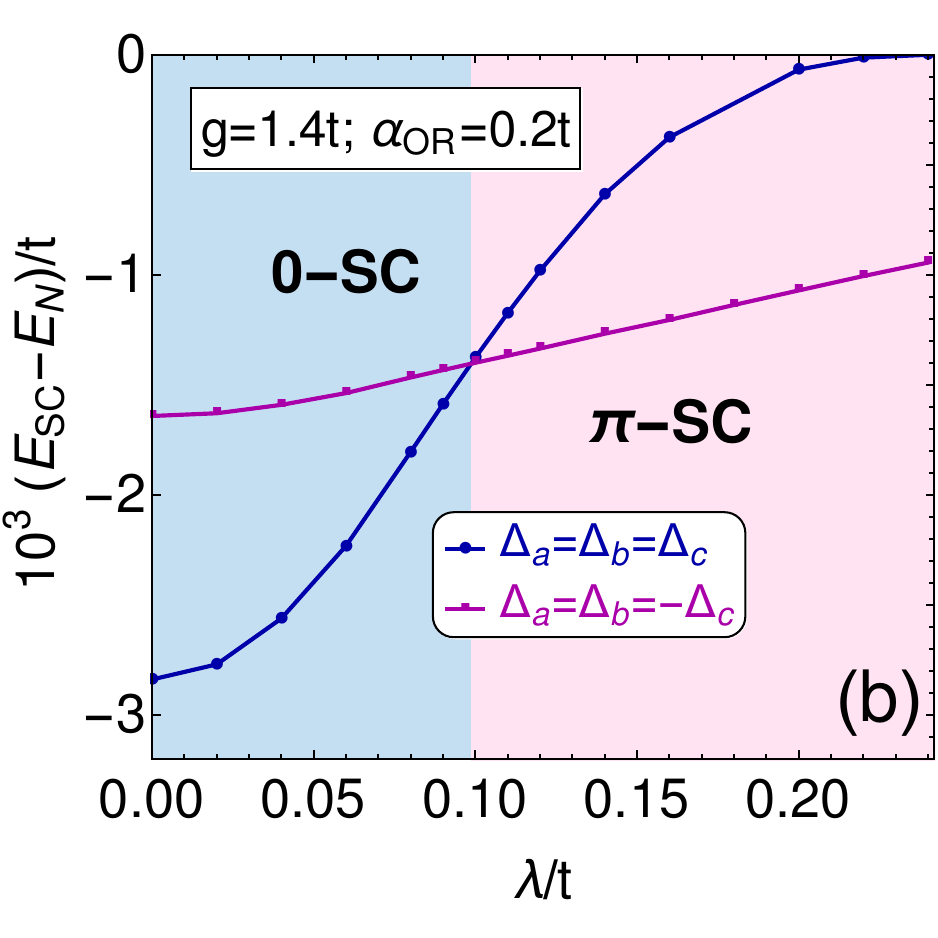}
\includegraphics[width=0.19\textwidth]{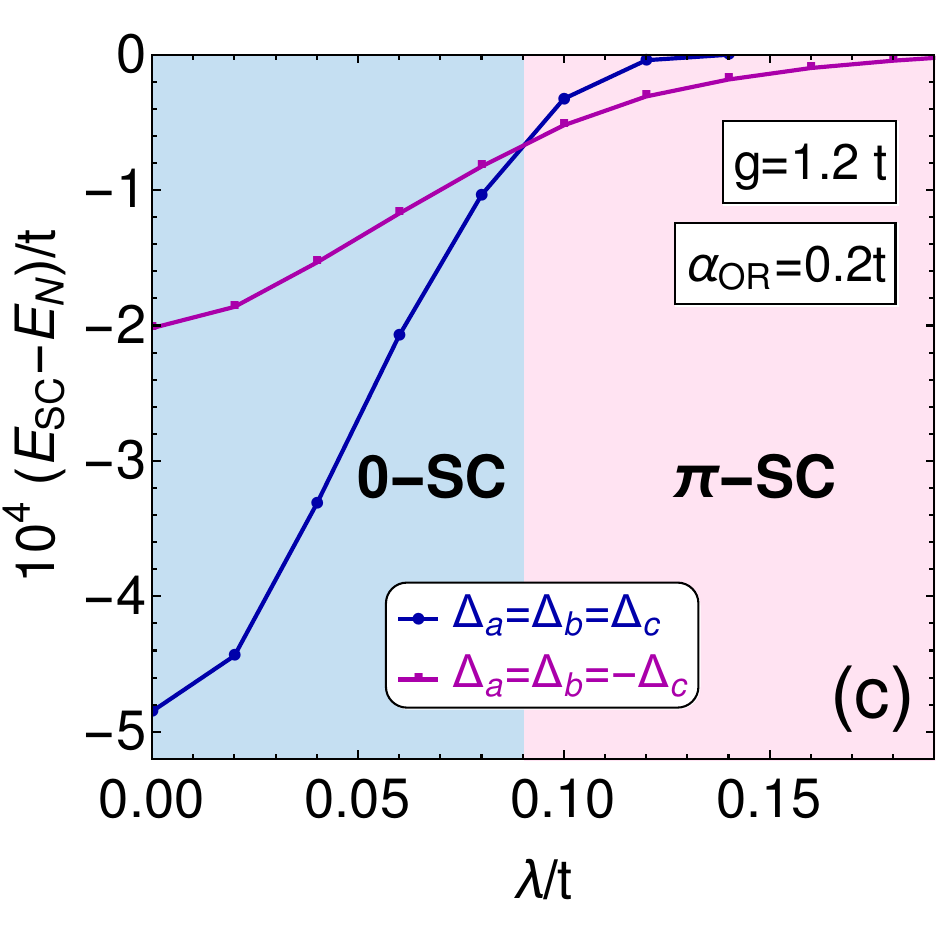}
\includegraphics[width=0.19\textwidth]{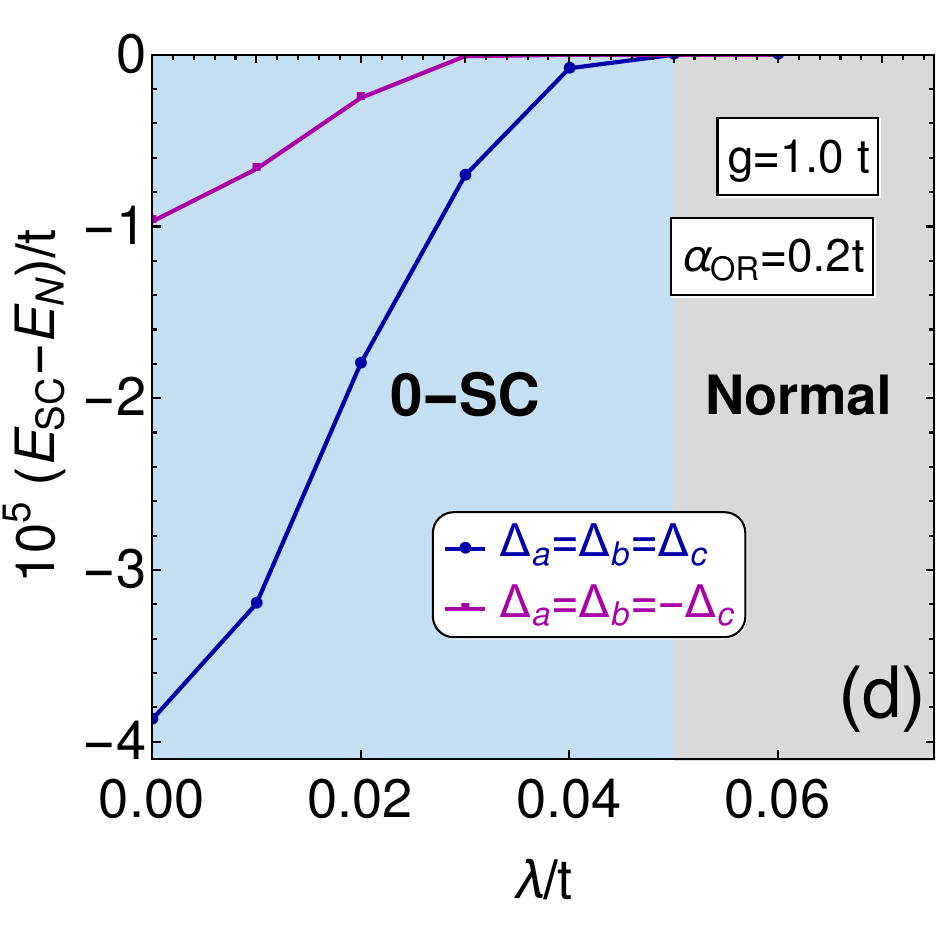}
\includegraphics[width=0.19\textwidth]{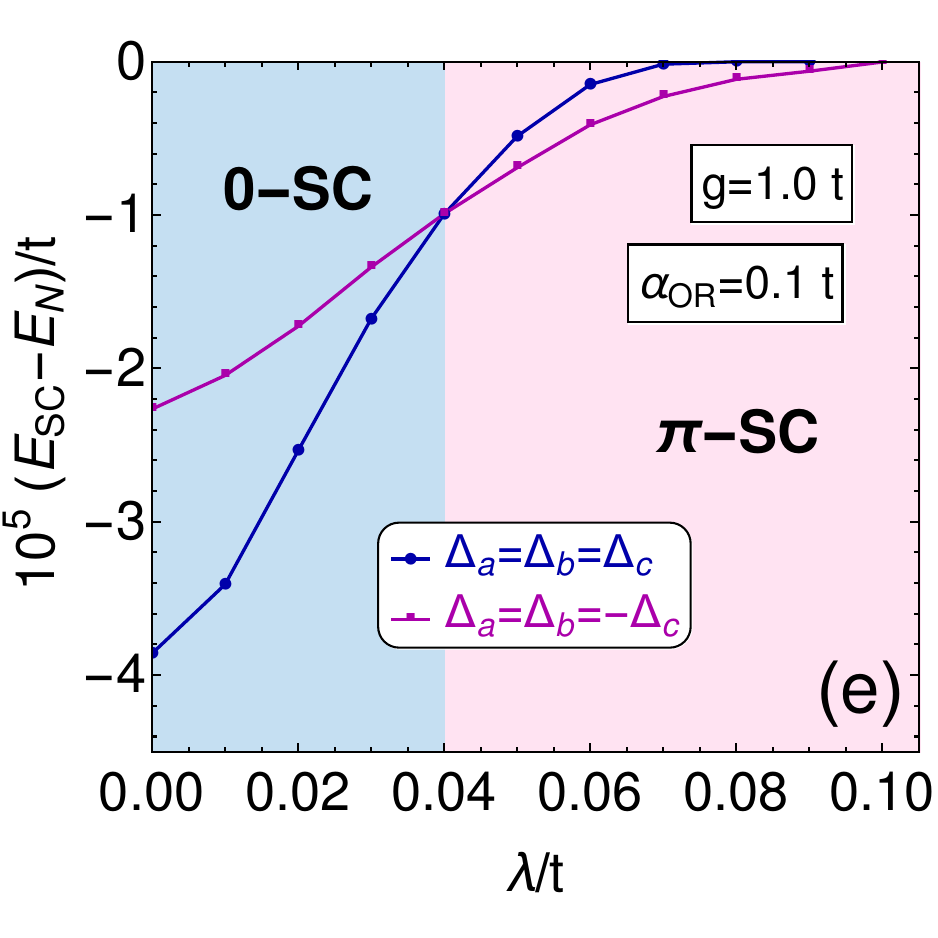}\\
\includegraphics[width=0.19\textwidth]{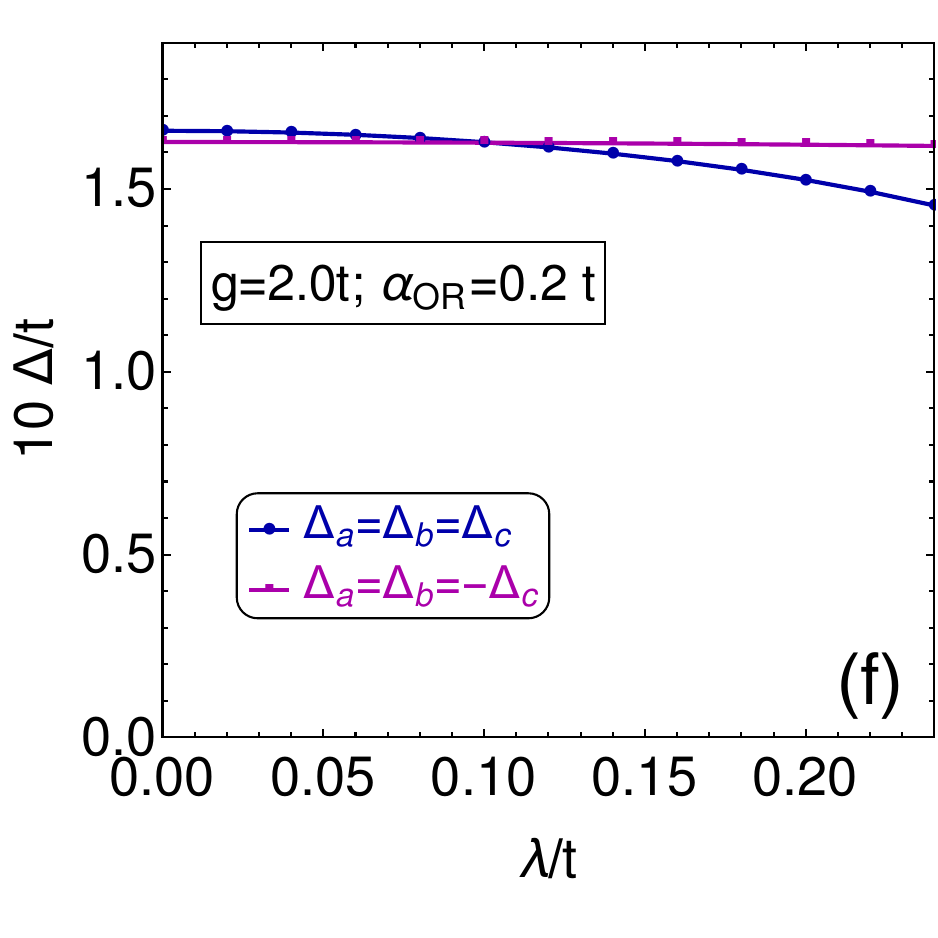}
\includegraphics[width=0.19\textwidth]{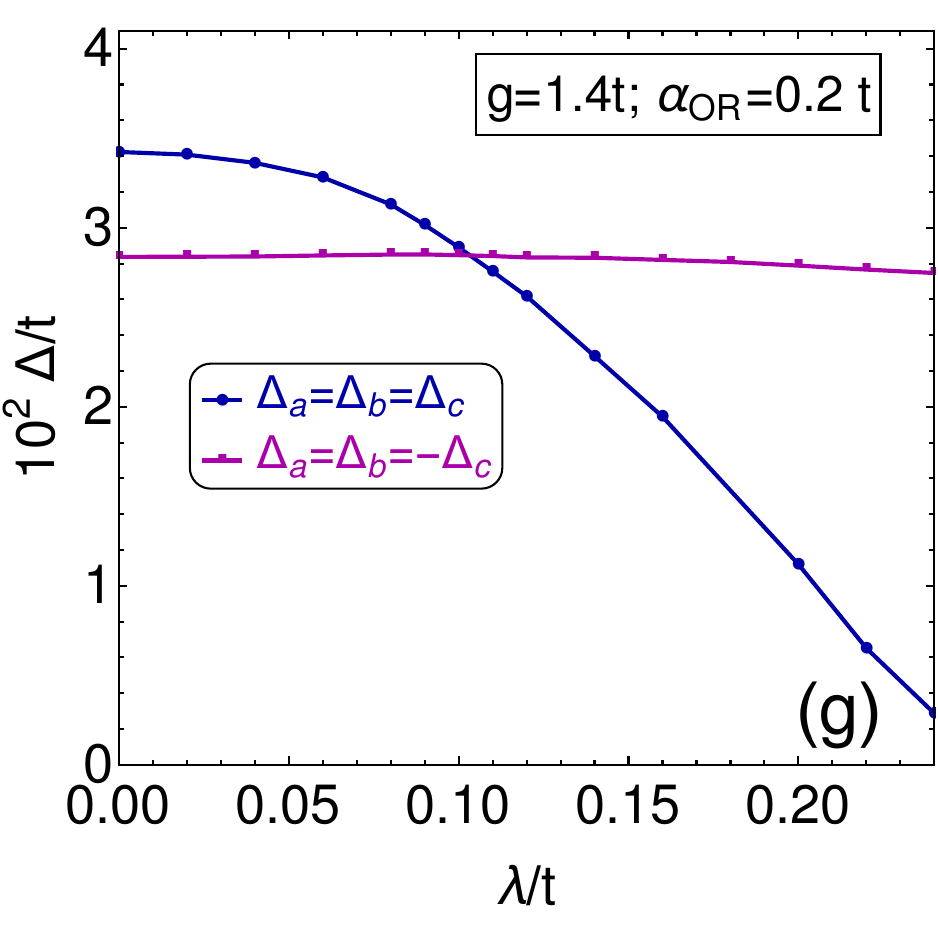}
\includegraphics[width=0.19\textwidth]{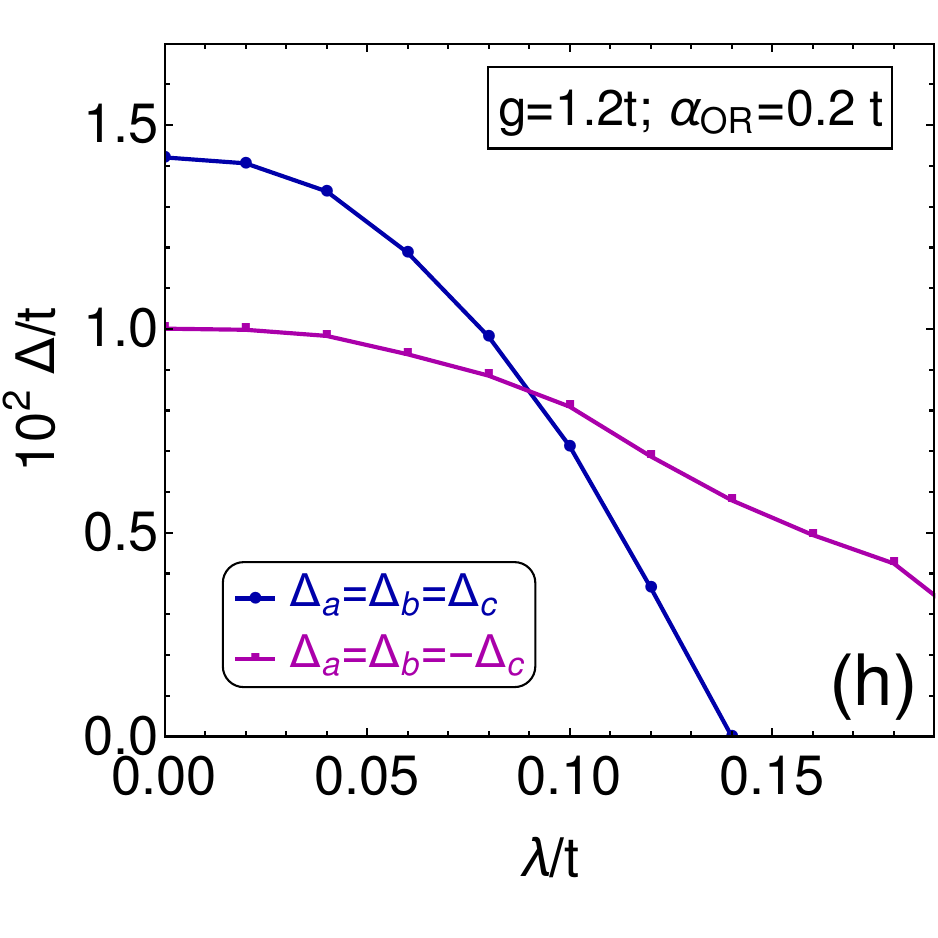}
\includegraphics[width=0.19\textwidth]{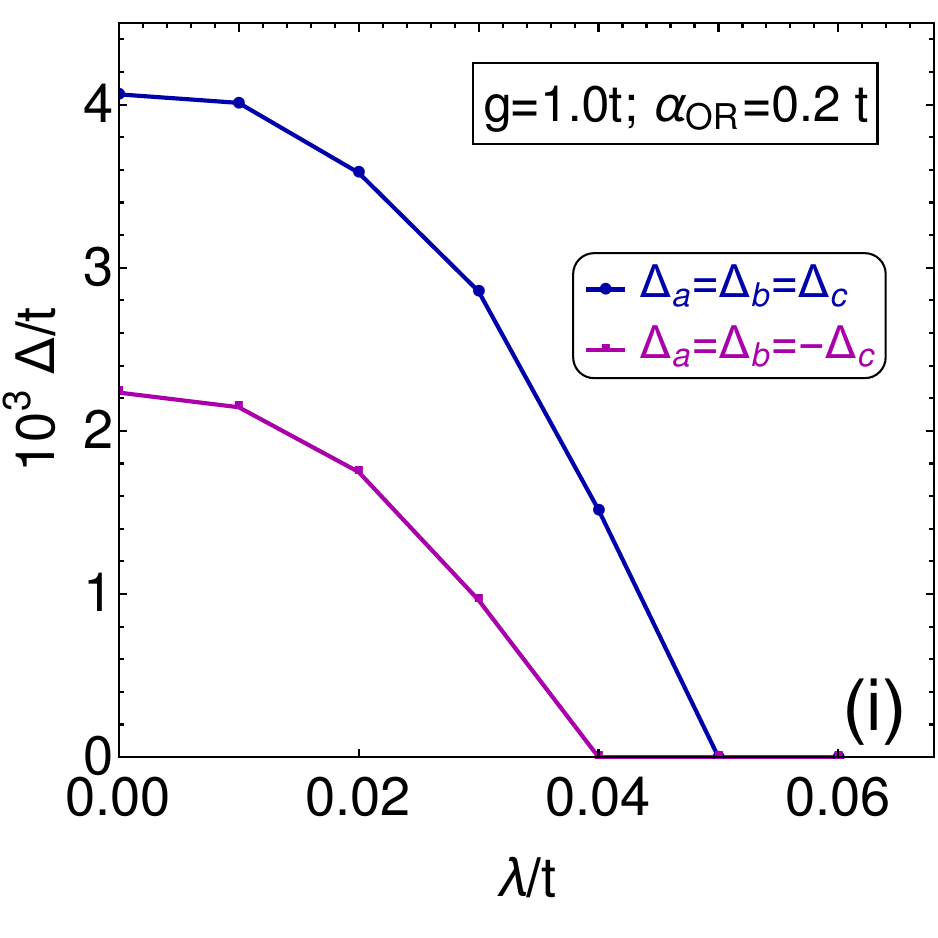}
\includegraphics[width=0.19\textwidth]{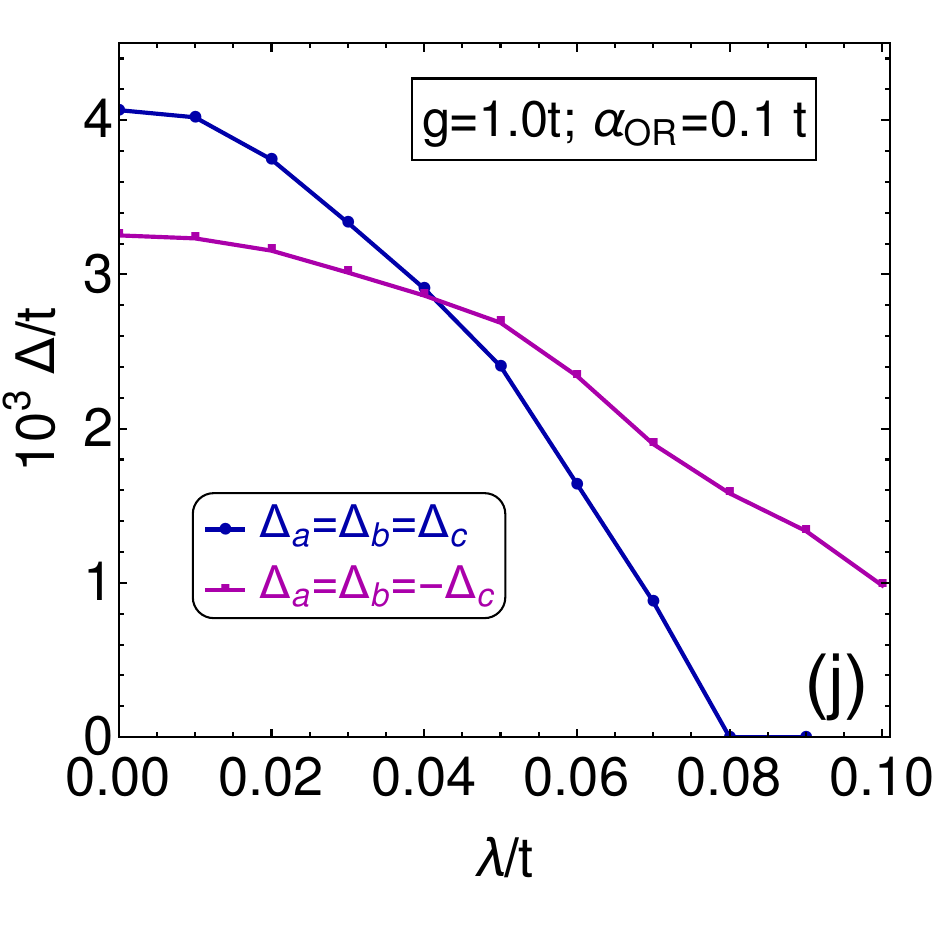}
\protect\caption{
(a-e) Plots of the superconducting free-energy as a function of the inversion-asymmetric interlayer parameter $\lambda$ for the two most relevant configurations (i.e. conventional (0-SC) blu lines and unconventional $(\pi$-SC) purple lines)
by considering different values of the pairing strength $g$. 
The orbital Rashba parameter has been fixed to $\alpha_{OR}=0.2t$ for figures(a-d), while it is $\alpha_{OR}=0.1t$ in panel (e). A system of $n_z=12$ layers has been considered and the other parameters are: $t_\perp=1.5t, \mu=-0.4 t, \eta=0.1$.
The crossing of the lines in (a-c) and (e) gives the transition value of the inversion asymmetric interlayer parameter $\lambda$. This critical value is almost unaffected by the change of $g$ when the same value of $\alpha_{OR}$ is considered. Namely  we have: $\lambda_c(g=1.2t)=0.0904t\simeq 0.09t$, $\lambda_c(g=1.4t)=0.0987t\simeq 0.10t$ and  $\lambda_c(g=2.0t)=0.1014t\simeq 0.10t$. 
For $g=1.0t$ and $\alpha_{OR}=0.2t$ the intermediate 0-$\pi$ transition is not anymore observed and the system undergoes a direct changeover into a normal state. 
Taking a smaller value of $\alpha_{OR}$ (see panel (e)), we recover again the intermediate 0-$\pi$ transition also for pairing strength $g=1.0\,t$. 
}
\label{fig:phasediagramvsg}
\end{figure*}

\begin{figure*}[bt]
\includegraphics[height=4.cm]{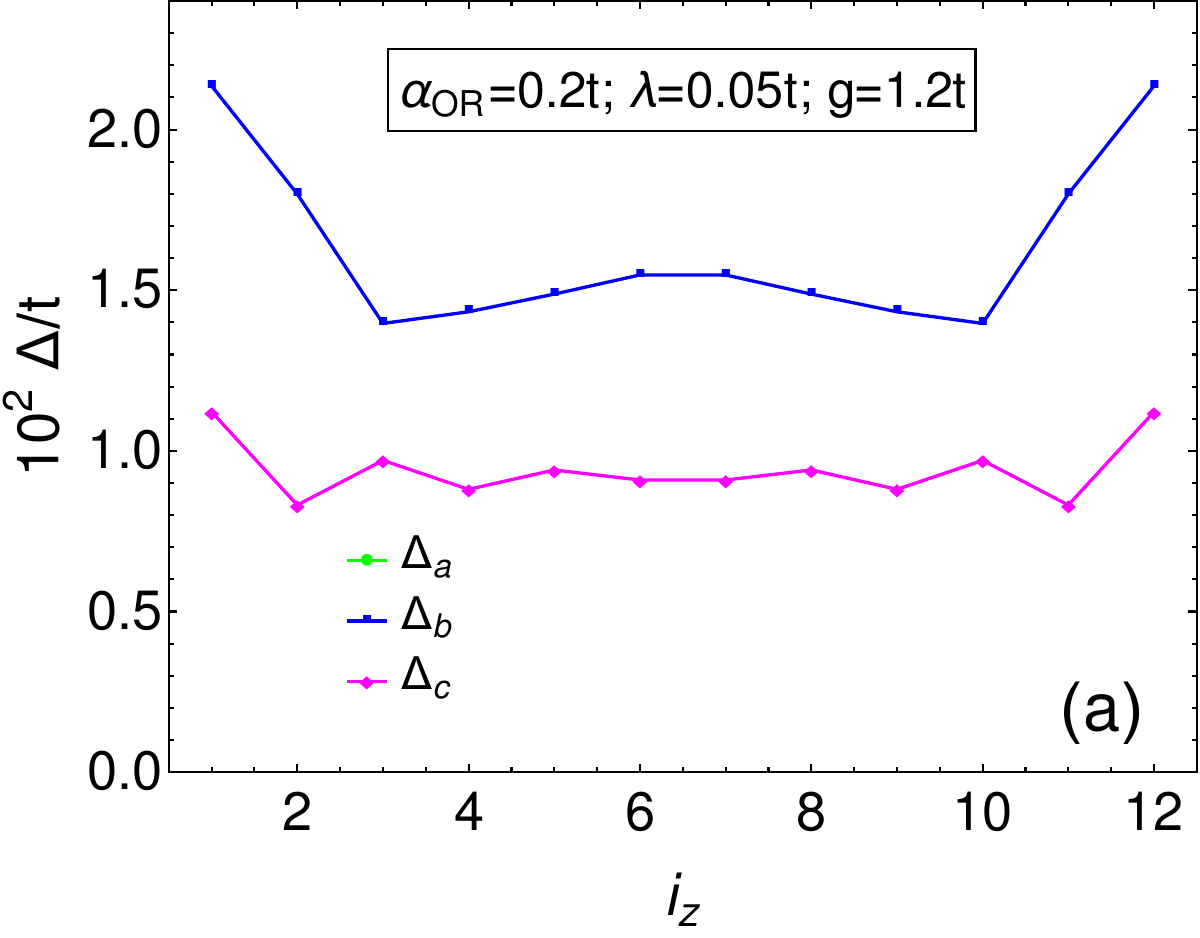}\hspace{0.5cm}
\includegraphics[height=4.cm]{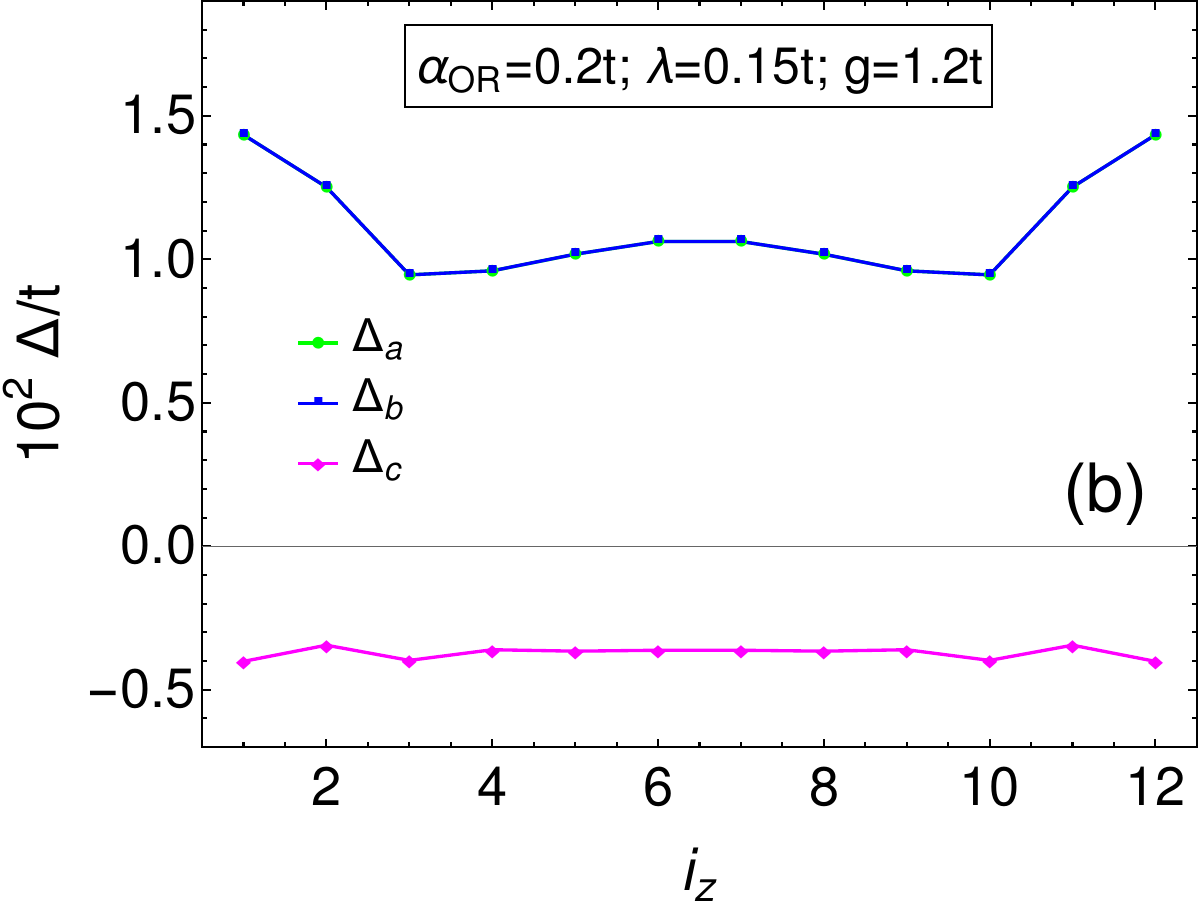}\hspace{0.5cm}
\includegraphics[height=4.cm]{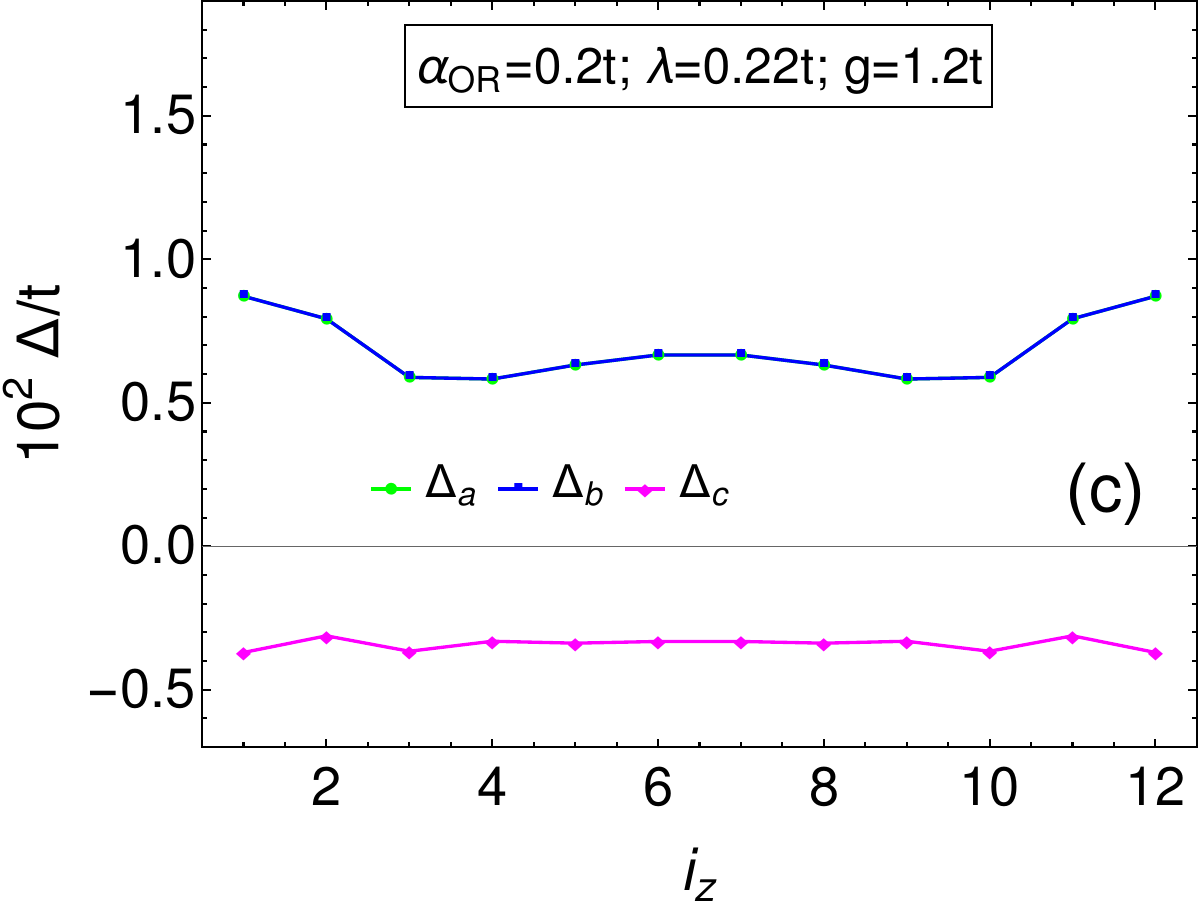}
\protect\caption{We report the spatial profile of the superconducting order parameter $\Delta_{\alpha} (\alpha=a,b,c)$ along the $z$-direction for a slab with $n_z=12$ layers,  pairing strength $g=1.2\,t$, and orbital Rashba coupling $\alpha=0.2\,t$. The OP has been calculated self-consistently described in the main text.
In panel (a) for $\lambda=0.05\,t$ we find that all the three components of the OP have the same sign (hence the system is in the conventional $0-$SC phase), in (b) and (c) for $\lambda=0.15\,t$ and $\lambda=0.22\,t$, $\Delta_c$ has opposite sign with respect to $\Delta_{a,b}$ thus realizing the unconventional $\pi-$SC phase).}
\label{fig:selfcons}
\end{figure*}

We have followed the evolution of the phase diagram to understand the role of the superconducting pairing strength. In Fig. \ref{fig:phasediagramvsg} we report the overall effect of the pairing strength going from $g/t=2.0$ to $g/t=1.0$ as a function of $\lambda$ for a pair of representative values for the orbital Rashba coupling $\alpha_{OR}$. We find that the critical $\lambda$ to induce the $0$-$\pi$ transition is practically unaffected when the pairing coupling $g$ is varied from $2\,t$ to $1.2\,t$ (Fig. \ref{fig:phasediagramvsg} (a)-(c)). On the other hand, for $g/t=1.0$ we have that the transition from 0- to $\pi$-phase does not occur and a change in the inter-layer $\lambda$ coupling directly brings the superconducting into the normal state at $\alpha_{OR}=0.2\,t$. However, if one assumes that the orbital Rashba coupling is scaled to $\alpha_{OR}=0.1\,t$ than one recovers the 0-$\pi$ phase transition as demonstrated in Fig. \ref{fig:phasediagramvsg} (e). This result clearly indicates that the potential to drive the superconducting phase into a $\pi$- or normal state is a robust effect and that the relative ratio between the intra- and inter-layer asymmetric interactions can set out whether the 0-normal phase transition is obtained with an intermediate $\pi$-phase or without passing through this state.
Finally, we show that such delicate interplay between the 0-, $\pi$- and normal phases is also imprinted into the evolution of the superconducting order parameters as reported in Fig. \ref{fig:phasediagramvsg} (f)-(j).  
For completeness, we have also demonstrated that the 0-$\pi$ transition can be obtained at $g/t=1.2$ within a self-consistent analysis that is able to capture the non-uniform spatial dependence of the order-parameter along the $z$-direction (Fig. \ref{fig:selfcons}). 

\subsection{Role of inter-orbital mixing for the single-particle electronic states}

\begin{figure}[bt]
\includegraphics[width=0.48\columnwidth]{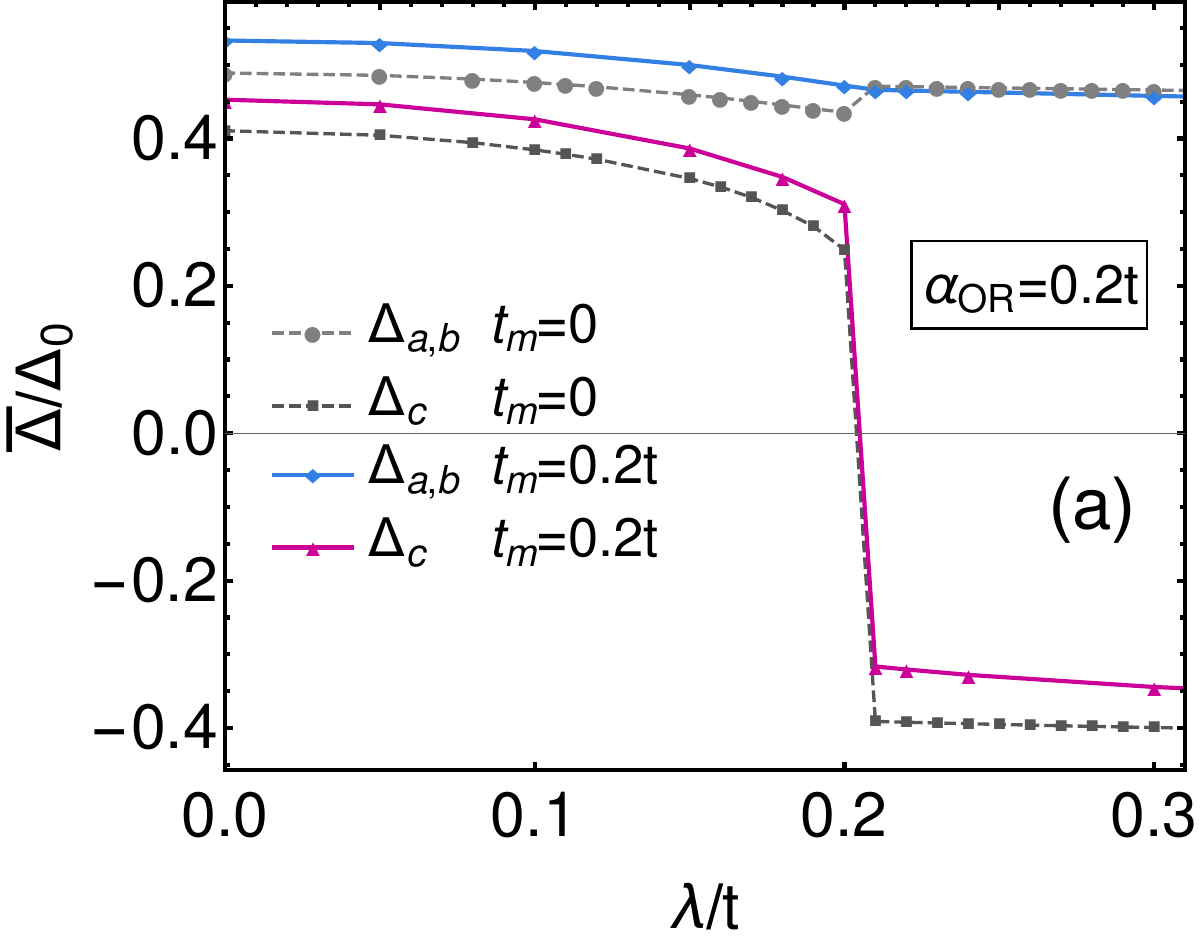} \hspace{0.1cm}
\includegraphics[width=0.47\columnwidth]{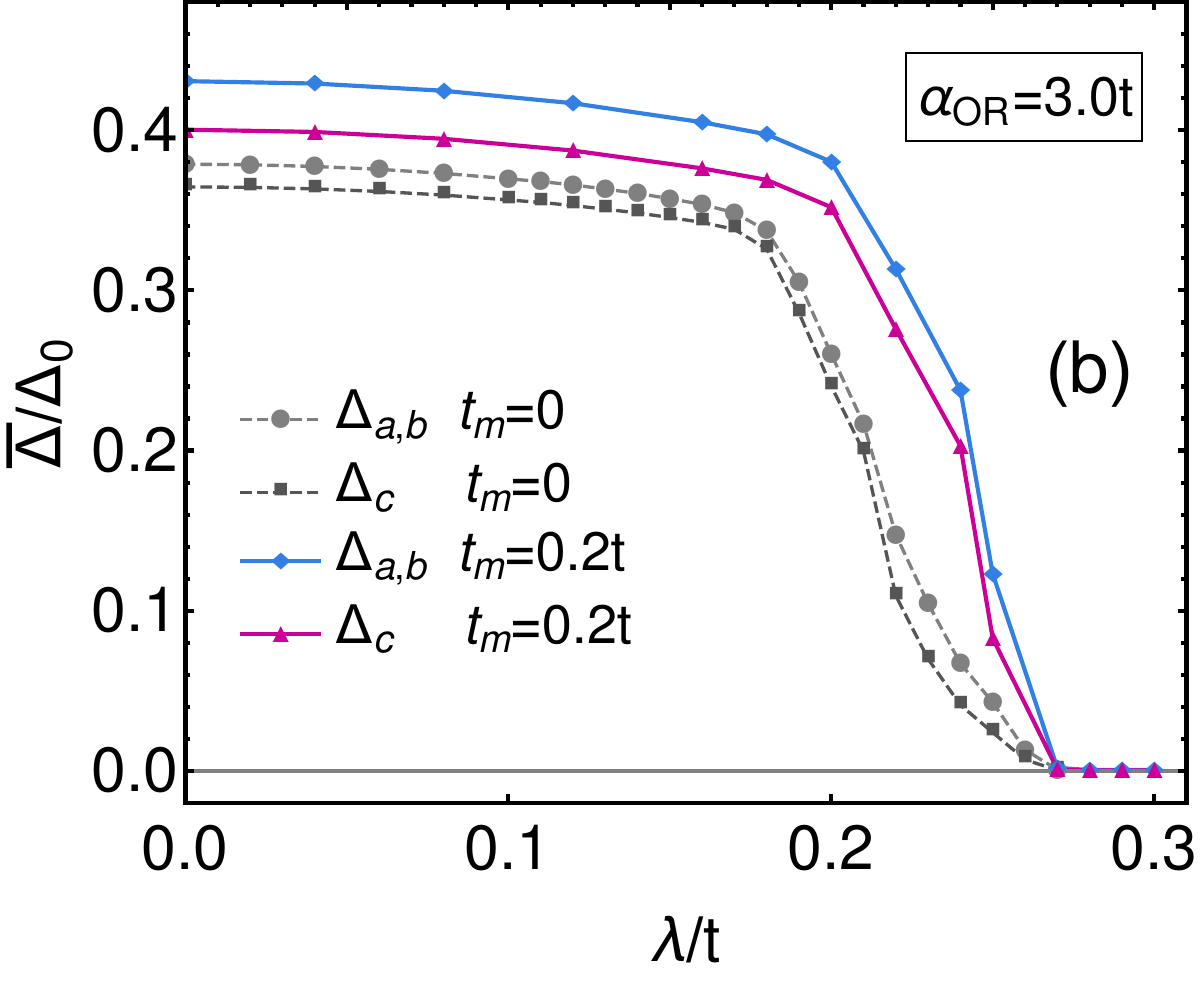} 
\protect\caption{We compare the results obtained in Figs. 2 b-c (which are here reproduced as gray lines) with those obtained adding {\it orbital mixing} hopping terms. We see that the 0-SC to $\pi$-SC transition and the 0-SC to normal state one are unaffected by the inclusion of orbital mixing terms. Specifically, we show the behavior of the order parameter in the inner
side of the system $\bar{\Delta}$ as function of the surface interlayer coupling $\lambda$ in the regimes of
weak (a) and strong (b) orbital Rashba interaction, namely $\alpha_{OR} = 0.2\,t$ and $\alpha_{OR} = 3.0\,t$, respectively. In both cases the critical value of $\lambda$ is not changed with the inclusion of an orbital mixing hopping term $t_m=t_{m\perp}=0.2t$, the other parameters are as in Fig.2 b-c of main paper, i.e. $n_z=6, t_\perp=1.5t, \mu=-0.4\,t, \eta=0.1$,and $\beta=0.1$.}
\label{fig:orbitalmix}
\end{figure}

We point out that the employed tight-binding electronic structure has realistic features if one considers that the bands at the Fermi level are formed out of anisotropic atomic orbitals of $p$ or $d$ type for instance. Due to symmetry arguments it is known that in a cubic or tetragonal environment the $(d_{xy},d_{xz},d_{yz})$ orbitals belonging to the so-called $t_{2g}$ sector have only directional non-vanishing nearest-neighbor hopping amplitudes.
Within a tight-binding formulation of the electronic structure one can apply the Slater-Koster rules \cite{Slater1954} and determine the allowed hopping amplitude between Wannier configurations on different atoms whose distance is parameterized in terms of the bond angle. This approach yields that, for instance, $d_{xy}$ atomic state can hybridize only with $d_{xy}$ configurations in the $x-y$ plane along the [100] and [010] cubic directions and similarly for the other orbitals. Thus, it is also suited for elemental materials like Ti, V, Nb, etc., and it can also apply to more complex metals as those occurring in the realm of transition metal oxides. 
\\
Apart from these general considerations, since distortions would lead to deviations from ideal electronic structure above discussed, we have included extra terms in the single particle part of the Hamiltonian which lead to mixing of orbitals along the symmetry direction. This analysis has been performed to further investigate the role of the orbital mixing on the phase diagram. 
   
Additional terms in the Hamiltonian are:
(1) intra-layer hopping terms, 
\be 
{\cal H}_{m,\parallel}= \frac 1N \sum_{i_z=1}^{n_z} \sum_{\bk} \sum_{\alpha,\beta} \sum_{\sigma=\up,\dw}
 d^\dagger_{\alpha,\sigma}(\bk,i_z) \varepsilon_{\alpha\beta}(\bk) d_{\beta,\sigma}(\bk,i_z)
\ee
where the diagonal terms ($\varepsilon_{\alpha\alpha}\equiv \varepsilon_\alpha$) are those of Sect. II, and
\bea
\varepsilon_{ab}(\bk)&=&-2 t_m \beta (\cos(k_x) + \cos(k_y)) \nonumber\\ 
\varepsilon_{ac}(\bk)&=&-2 t_m (\cos(k_x) + \beta \cos(k_y)) \nonumber \\
\varepsilon_{bc}(\bk)&=&-2 t_m (\beta \cos(k_x) + \cos(k_y))\nonumber
\eea
(2) inter-layer hopping terms
\be 
{\cal H}_{m,\perp}= -\frac 1N 
\sum_{\langle i_z, j_z \rangle} \sum_{\bk} \sum_{\alpha,\beta} \sum_{\sigma=\up,\dw}
d^\dagger_{\alpha,\sigma}(\bk,i_z) t^{\alpha\beta}_\perp d_{\beta,\sigma}(\bk,j_z)
\ee
where $t^{\alpha\beta}_\perp=t_\perp$ when $\alpha=\beta$ and $t^{\alpha\beta}_\perp=t_{m,\perp}$ when $\alpha \neq \beta$ ($\alpha,\beta=a,b,c$), and $\langle ... \rangle$ is restricted to adjacent layers.

The results are reported in Fig. \ref{fig:orbitalmix} for a representative case of $t_{m}=0.2 t$ and two values of the orbital Rashba coupling $\alpha_{OR}$ which allow to drive the superconductor into the $\pi$-phase and into the normal metal state as a function of the inter-layer interaction $\lambda$. As one can see, the effects of the inter-orbital mixing are  negligible and the 0-$\pi$ or 0-Normal phase transitions occur at the same values of the $\lambda$ coupling as in the case with $t_{m}=0$. This analysis confirms that the phenomenology is robust to changes in the electronic structure.

\subsection{Inter-orbital pairing interaction}

Here, we consider the role of the inter-orbital pairing interaction. The aims are to assess whether the inter-orbital pairing influences the phase diagram and the potential link with the $\pi-$phase. The analysis has been performed with and without the inter-orbital hopping. Additionally, we follow a representative case of $\alpha_{OR}=0.2 t$ and scan the phase diagram for different values of $\lambda$. 

We start by pointing out that the inter-orbital pairing amplitude is expected to be anisotropic in the momentum space, due to the orbital Rashba terms, and to have a major role only nearby the points where Fermi lines with different orbital character cross each other. Another important aspect is that the mixing of the orbitals can arise both from the orbital Rashba couplings and from the inter-orbital hoppings with a different impact  on the inter-orbital pairing. On such basis, we have taken into account these aspects and analyzed the role of the inter-orbital pairing interaction on the phase diagram. 

\begin{figure}[bt]
\includegraphics[width=0.48\columnwidth]{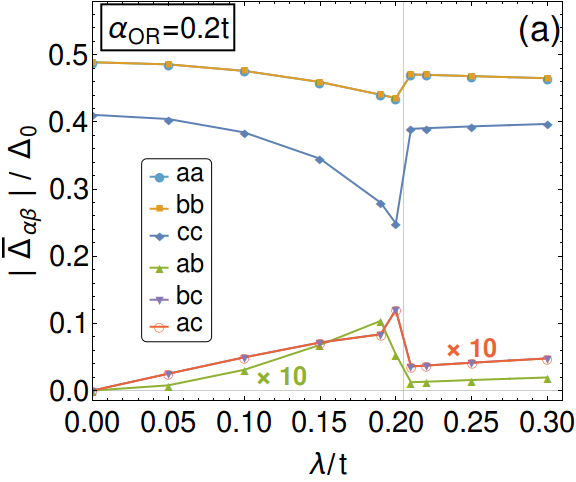} \hspace{0.1cm}
\includegraphics[width=0.49\columnwidth]{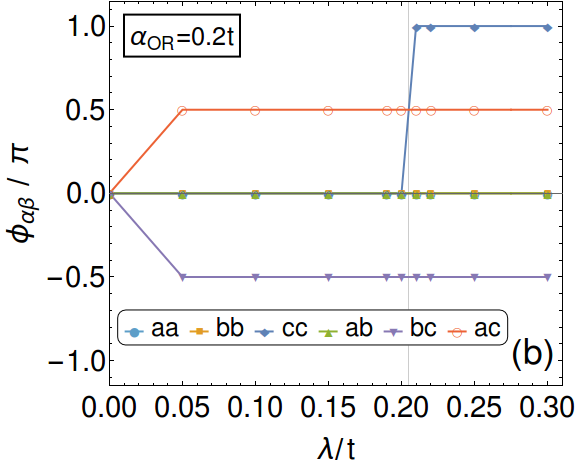} 
\protect\caption{Behavior of the superconducting order parameter components $\Delta_{\alpha\beta}=|\Delta_{\alpha\beta}| \exp(i \phi_{\alpha\beta})$ (with $\alpha,\beta=a,b,c$) [(a) amplitude and (b) phase] in the central layer ($i_z=n_z/2$)   as a function of the inter-layer asymmetric interaction $\lambda$ for a fixed value of the orbital Rashba coupling $\alpha_{OR}=0.2t$. The other parameters used are the same as in Fig.2(b) of main paper (i.e. $n_z=6, t_\perp=1.5\,t, \mu=-0.4t, \eta=0.1$),  with the additional inclusion of the inter-orbital pairing interaction $g_{od}$. 
The amplitude of the inter-orbital OP have been multiplied by a factor of 10 to be visible in the same plot. $\Delta_{aa}, \Delta_{bb}$ and $\Delta_{ab}$ are real and positive, thus their phases are zero for any $\lambda$.  We stress that the values of the diagonal OP, $\Delta_{\alpha\alpha}\equiv\Delta_{\alpha}$ (with $\alpha=a,b,c$), do not change appreciably, if compared with Fig.2(b) of main text, in the self-consistent evaluation upon the inclusion of the interorbital pairing, hence the onset of the $\pi$-phase remain unchanged. We point out that here the inter-orbital hopping amplitude is zero.}
\label{fig1-inter-g}
\end{figure}

\begin{figure*}[bt]
\includegraphics[height=3.5cm]{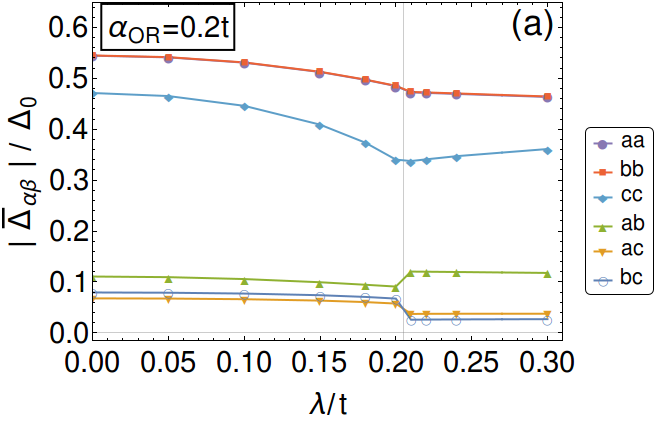} \hspace{0.4cm}
\includegraphics[height=3.5cm]{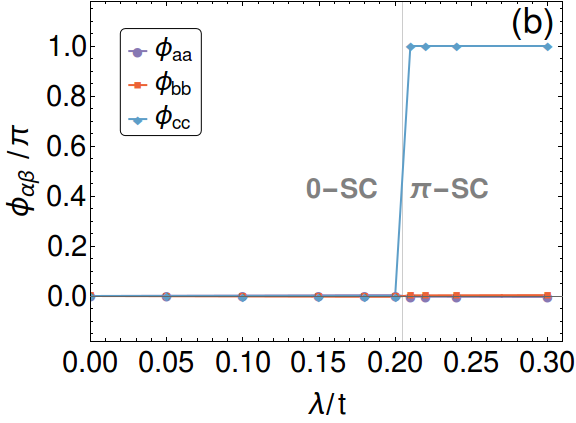} 
\includegraphics[height=3.5cm]{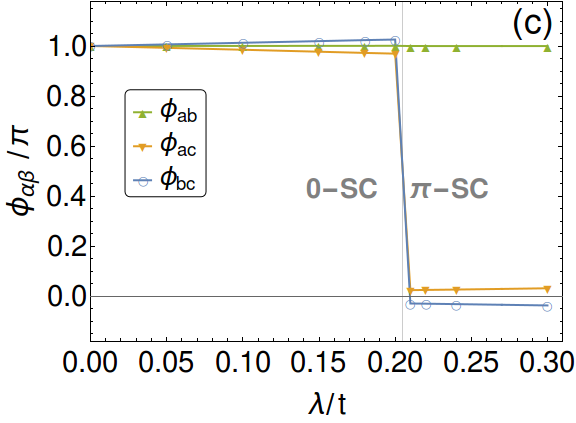} 
\protect\caption{Behavior of the superconducting order parameter components $\Delta_{\alpha\beta}=|\Delta_{\alpha\beta}| \exp(i \phi_{\alpha\beta})$ (with $\alpha,\beta=a,b,c$) [(a) amplitude and (b-c) phase of the OP] in the central layer ($i_z=n_z/2$)   as a function of the inter-layer asymmetric interaction $\lambda$ for fixed value of the orbital Rashba coupling $\alpha_{OR}=0.2t$. In this analysis orbital mixing hoppings and  interorbital pairing are included. The value of the interorbital hopping parameter here used is $t_m=t_{m\perp}=0.2t$, while other parameters are the same as in Fig.2(b) of main paper (i.e. $n_z=6, t_\perp=1.5t, \mu=-0.4t, \eta=0.1$). 
It is evident that, even if the amplitudes of the several components have  changed from the case with absence of inter-orbital hoppings and pairing, a $\pi$-phase it is still present and its onset occurs at the same value of the parameter $\lambda$. }
\label{fig2-inter-g}
\end{figure*}

The overall outcome is quite clear. Firstly, we find that the presence of an inter-orbital pairing interaction does not affect the character and the structure of the phase diagram. This is confirmed by the fact that the critical $\lambda$ for the transition into the $\pi$-phase is substantially unaffected by the presence of the inter-orbital order parameters (Figs. \ref{fig1-inter-g} -\ref{fig2-inter-g}). 

It is instructive to start considering the nature of the inter-orbital pairing for the case of vanishing inter-orbital mixing in the single particle spectrum. Indeed, for such physical circumstance, we remark that non-vanishing $\Delta_{ac}$ and $\Delta_{bc}$ occur only when the inter-layer $\lambda$ term is non zero and the order parameters have always a $\pi$-phase difference (Fig. \ref{fig1-inter-g}). This behavior clearly indicates that the $\lambda$ term tends to favour a phase difference between the inter-orbital order parameters that are mainly inolved in the $\pi$-phase. Thus, the $\lambda$ coupling shapes the pair correlations to drive an orbital-dependent phase rearrangement of the superconducting state. 

For this physical case, it is also interesting to touch on symmetry aspects behind the fact that the order parameters $\Delta_{ac}$ and $\Delta_{bc}$ develop a $\pi$-phase difference. We argue that their behavior reflects the symmetry properties of the $\lambda$ term. Since $\lambda$ coupling breaks the mirror symmetries with respect to the $xz$ and $yz$ planes $\Delta_{ac}$ and $\Delta_{bc}$ have to be non-vanishing. However, we argue that, due to the preservation of one of the mirror symmetry with respect to the diagonal in the $xy$ plane, the superposition of the order parameters can be conserved thus favoring a $\pi$-phase difference (i.e. their combination cancels out). Hence, we also argue that the inter-orbital pair correlations act like a seed for inducing a phase rearrangement in the intra-band superconducting order parameter that optimally lowers the energy.  

Finally, as demonstrated in Fig. \ref{fig2-inter-g}, the inclusion of the inter-orbital hoppings indicates that the inter-band $\pi$-phase difference do not occur at small $\lambda$ and one needs to overcome a critical threshold for the $\lambda$ coupling to stabilize a complete orbital reconstruction of the superconducting state (Fig. \ref{fig2-inter-g} (b)-(c)) that indeed corresponds to the identified $\pi$-phase in the phase diagram.

\subsection{Effects of interlayer hopping}

\begin{figure}[bth]
\begin{center}
\includegraphics[height=2.65cm]{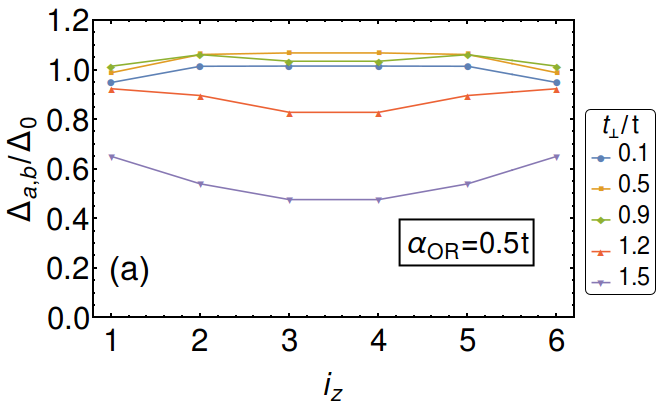}
\includegraphics[height=2.65cm]{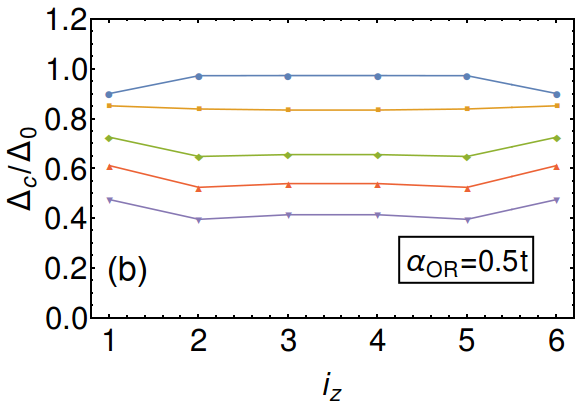} \\
\vspace{0.3cm}
\includegraphics[height=2.65cm]{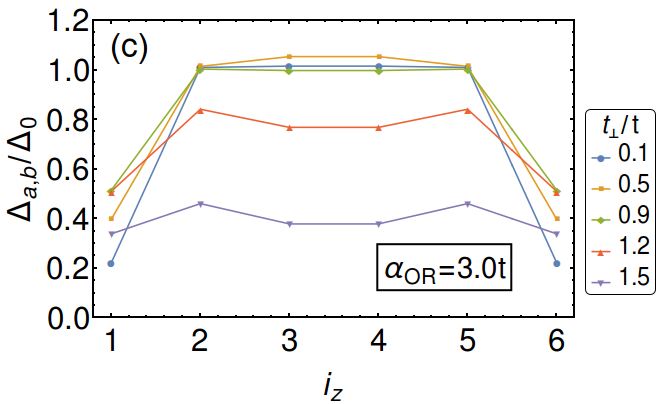}
\includegraphics[height=2.65cm]{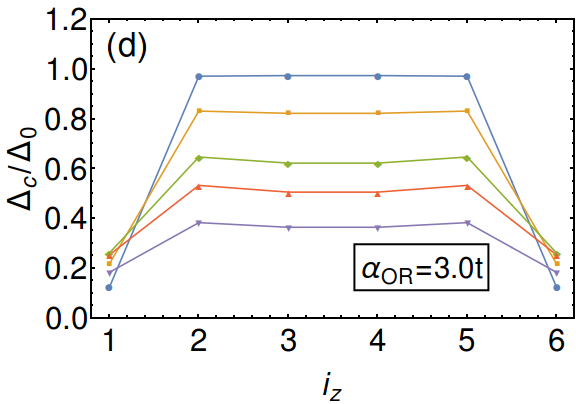} 
\end{center}
\protect\caption{Spatial profile of the superconducting order parameter $\Delta_\alpha (\alpha=a,b,c)$ along the $z$ direction at $n_z=6$ for five different values of $t_\perp$, in the case of weak (panels (a) and (b)) and strong (panels (c) and (d)) orbital Rashba coupling. The surface interlayer interaction $\lambda$ is set to zero.} 
\label{figTP}
\end{figure}

Here, we analyze the influence of the interlayer hopping $t_\perp$ on the order parameter in the superconducting phase. In Fig.~\ref{figTP} we show the profile of $\Delta_\alpha(i_z)$ along the $z$ direction for a superconductor with $n_z=6$, considering two different values of OR coupling and several values of $t_\perp$, in absence of the surface interlayer interaction $\lambda$. 

\begin{figure}[th]
\begin{center}
\includegraphics[height=2.65cm]{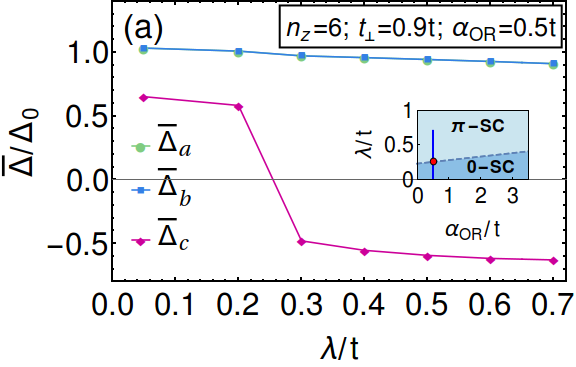}
\includegraphics[height=2.65cm]{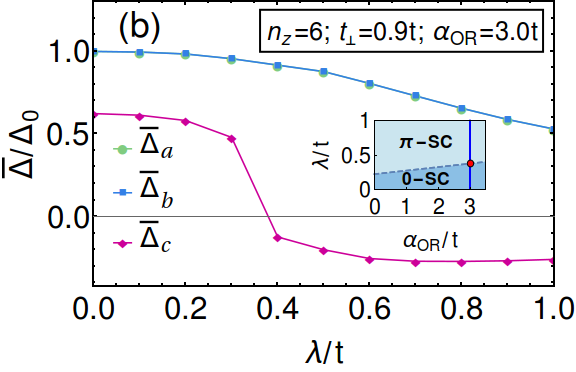} \\
\vspace{0.3cm}
\includegraphics[height=2.65cm]{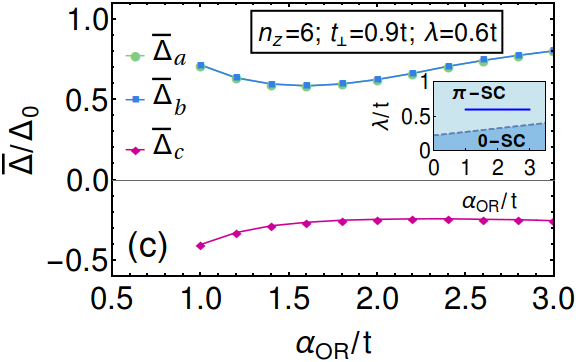}
\includegraphics[height=2.65cm]{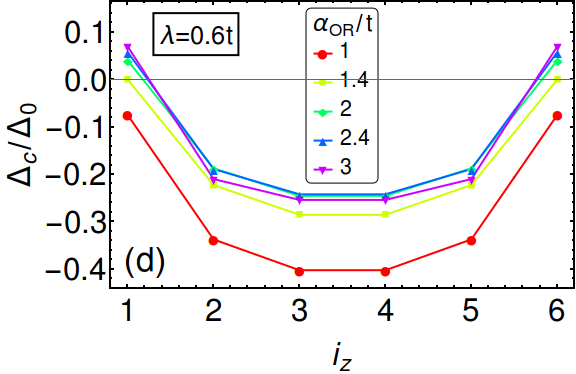} 
\end{center}
\protect\caption{(a)-(c) Behavior of the order parameter $\bar{\Delta}_\alpha$ inside the system (i.e. in the central layer $i_z=3$)  for $t_\perp=0.9t$. In the insets a schematic phase diagram in the $(\alpha_{OR},\lambda)$-plane is shown, where the blu line mark the path followed in the figure and the red dot denotes the transition point. Specifically, in (a) we present the results for $\alpha_{OR}=0.5t$, in (b) for  $\alpha_{OR}=3.0t$, while in (c) we follow a horizontal path in the ($\alpha_{OR},\lambda$)-plane, with $\lambda=0.6t$ varying $\alpha_{OR}$. Finally, in (d) we report the spatial profile of $\Delta_c$ along the $z$-direction corresponding to the analysis performed in panel (c). Here we see that while  $\bar{\Delta}_c$ is always negative, $\Delta_c$ becomes positive in the outer layers for $\alpha_{OR}\gtrsim 1.3t$.} 
\label{figTP2}
\end{figure}

The effects of $t_\perp$ can be relevant and indeed the phase diagram reported in the Fig.2~(a) of the main text gets modified. A large amplitude of $t_\perp$ with respect to $t$ can destroy the superconducting state, even for small values of $\alpha_{OR}$ and $\lambda$. On the other hand, in the opposite regime of small $t_\perp$ one needs a significantly large amplification of $\lambda$  to get into the normal state. For this circumstance, one can typically obtain only 0-$\pi$ superconducting transition. In Fig.~\ref{figTP2}(a)-(c) we show the behavior of the order parameter in the inner side the system $\bar{\Delta}_\alpha$ (i.e. in the central layer $i_z=n_z/2$) for $t_\perp=0.9t$. We see that both for weak and strong values of the OR interaction the $0-\pi$ SC transition can be achieved. 

It is plausible to expect that the effective coherence length along $\hat{z}$ is proportional to the out-of-plane Fermi velocity, and thus one can argue that it scales with the amplitude of the inter-layer hopping. Such observation implies that the size of the SC is relevant for observing the SFE. Since the reduction of $t_{\perp}$ can alter the phase diagram with the normal state region being replaced by the $\pi$-SC configuration for the same strength of applied electric field, we predict that the inter-layer kinetic energy can be a suitable parameter to control the electric field effects on the superconductivity.

\section{Conclusions and discussion}  
We have demonstrated that by electrically tuning the surface orbital-polarization one can control both amplitude and phase of the superconductor.   
We have explicitly derived the microscopic origin of the surface couplings as due to the electrostatic potential. The induced interactions generally drive a complete reconstruction of the superconducting state with inter-band $\pi-$phase as well as superconducting-normal metal transition. The $0$-$\pi$ phase change is mainly first-order like, while the transition from superconductor to normal metal has weakly first-order precursors of the OP before it continously goes to zero. 
Concerning the $\pi$-phase, we expect that the sign frustration leads to an anomalous Josephson coupling in the case of inhomogeneous thin films.
Indeed, in the presence of non-magnetic disorder, the inter-orbital scattering between bands having opposite sign in the superconducting OP will result into a cancellation of the supercurrents and a behavior of an unconventional metal. Evidences of this state can be directly observed by phase sensitive superconducting interferometry \cite{Paolucci2019connecting}. Remarkably, the $\pi$-phase is compatible with the magnetic field dependence of the critical electric field that identifies the tansition from the superconducting state to a phase with vanishing critical supercurrent \cite{bours}.
\\
The obtained phase transitions are also linked to the character of the electron itinerancy of the superconducting thin film and, consequently, to its thickness. We have verified that the EF is more effective in a regime where the inter-layer kinetic energy is comparable to the planar one. Furthermore, the energy scales of the inversion asymmetric potentials at the surface for achieving the transitions are comparable to the bare hopping. This observation sets a clear reference for the electrical and orbital tunability of the superconducting phase. We point out that, since $\lambda$ and $\alpha_{OR}$ are proportional to the EF, with $t_{||} \sim 100$ meV, we predict that an electric field $\sim$ 30 mV/$\AA$ would suffice to observe the superconducting phase transitions, which is in the range of the experimental observations \cite{DeSimoniNatNano2018, PaolucciNanoLett2018, PaolucciPhysRevAppl2019, DeSimoni2019mesoscopic, Paolucci2019connecting}. We prove that the proportionality factor is a function of inter-atomic distances and of the distortions/strains at the surface.
Our findings thus indicate relevant paths for designing devices with electrically tunable superconducting orbitronics effects. In particular, central of our proposal is that the bands at the Fermi level can develop a non-vanishing orbital momentum, a fact that is ubiquituous in SCs with $p$- and $d$-bands at the Fermi level. Along this line, we predict that heterostructures with few layers of strong strainable and orbitally polarizable materials deposited on the surface of conventional superconductors would magnify the EF effects.

\begin{acknowledgments}
FG acknowledges the European Research Council under the European
Union’s Seventh Framework Programme (COMANCHE; European Research Council Grant No. 615187) and Horizon 2020 and innovation
programme under grant agreement No. 800923-SUPERTED.
\end{acknowledgments}

\appendix

\section{Microscopic derivation of the interactions induced by the surface electric field}

\begin{figure*}[bt]
\includegraphics[height=4.cm]{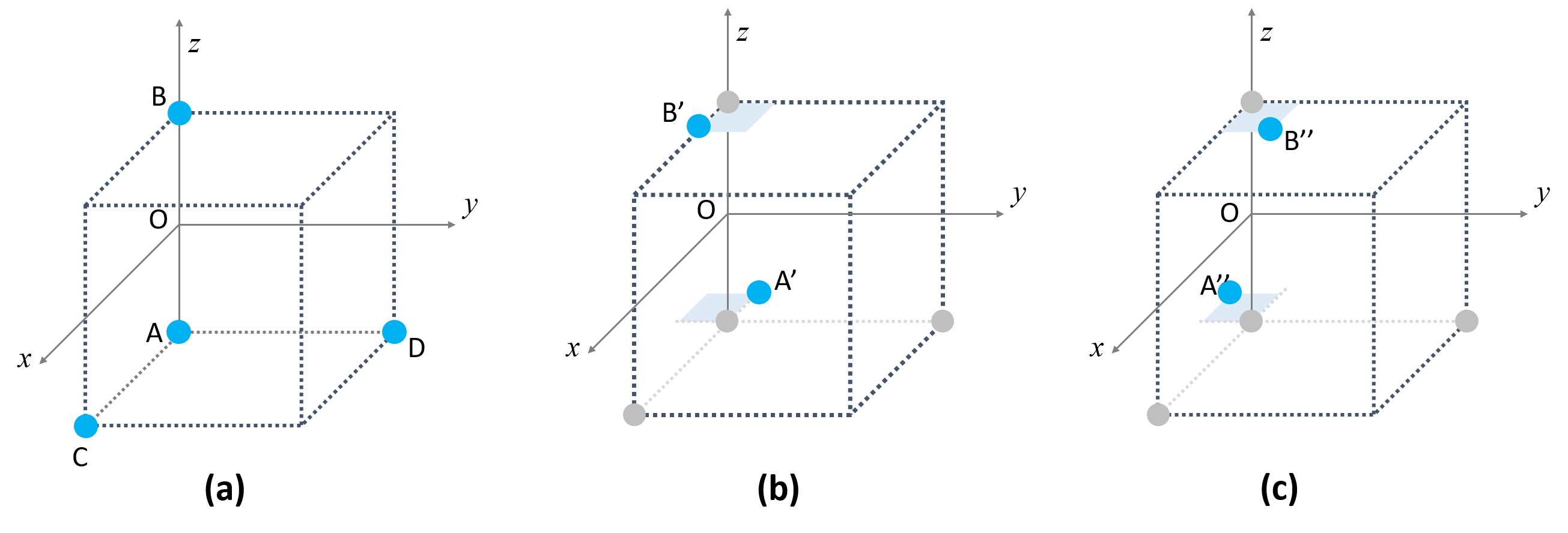}
\protect\caption{Schematic figure describing the atomic positions for the determination of the electrostatic energy associated to the intra- and inter-layer electronic processes. (a) sketch of the nearest neighbor atomic positions along the $(x,y,z)$ symmetry directions. (b) and (c) describe schematically the in-plane displacements that are related with the inter-layer orbital Rashba coupling.}
\label{fig:schema}
\end{figure*}
The external electric field on the surface of the superconductor is parallel to the $\hat{z}-$ direction and can be described by a potential $V_{s}=-E_{s} z$ with $E_s$ being constant in amplitude (assuming the electric charge $e$ is unit).  Following the approach that has been already applied to derive the surface orbital Rashba coupling \cite{Park2011,Park2012,Kim2013} we consider a Bloch state representation and explicitly evaluate the matrix elements of the electrostatic potential $V_s$. 
Since the translational symmetry is broken along the $z-$ direction, both for the finite thickness of the thin film and for the presence of the electric potential, the momentum is not a good quantum number. Thus, a representation with a Bloch wave function associated to each layer is suitable to evaluate the effects of the electric field and the way it enters in the tight-binding modelling.
Hence, we introduce the index $i_z$ to label different Bloch wave functions along the $z-$ direction as follows
\begin{eqnarray}
\psi_{{\bf k},\beta}({\bf r},i_z) =\frac{1}{\sqrt{N}}\sum_{\nu} \exp[i {\bf k} \cdot {\bf R}_{\nu, i_z}] \phi_{\beta}({\bf r}-{\bf R}_{\nu,i_z})
\end{eqnarray}
\noindent with the Bravais vector ${\bf R}_{\nu,i_z}$ identifying the position of the atoms in the $x-y$ plane for the layer labelled by $i_z$, $\beta$ indicating the atomic Wannier orbitals, and $N$ the total number of atomic sites. Here, it is central that the atomic Wannier functions span a manifold with non-vanishing angular momentum ${\bf L}$. 
To proceed further, we demonstrate how orbitally driven Rashba-like splitting occur in a $d$- (or equivalently $p$-) manifold restricting to the three-orbital subspace $\{d_{xy},d_{xz},d_{yz}\}$ (or $\{p_x,p_y,p_z\}$) due to the presence of the inversion symmetry breaking potential $V_s$ by evaluating the corresponding matrix elements for the above introduced Bloch states.
\\
For the derivation and the computation it is useful to introduce the following functions for the $d$-orbitals, for a given atomic position ${\bf R}_{\nu,i_z}$
\begin{eqnarray}
&&\phi_{xy}({\mathbf{r}})=f(r)\,xy\, \exp\left[-\frac{Z r}{n a_M}\right] \nonumber \\ 
&& \phi_{xz}(\mathbf{r})=f(r)\,xz\, \exp\left[-\frac{Z r}{n a_M}\right] \nonumber \\
&& \phi_{yz}(\mathbf{r})= f(r)\,yz\, \exp\left[-\frac{Z r}{n a_M}\right]
\end{eqnarray} 
\noindent with $f(r)=\frac14\sqrt{\frac{15}{\pi}} \; \left(-\left[\left(\frac{2 Z}{n a_M}\right)^3 \frac{(n-3)!}{2 n[(n+2)!]^3}\right]^{1/2}\right)
\left(\frac{2 Z}{n a_M}\right)^2 L_{n+2}^{5}(t)$, $Z$ being the atomic number, $n$ the principal quantum number, $t=2 Z r/(n a_M)$, $a_M=a_0(1+m_e/M)$ with $a_0$ the Bohr radius, $m_e$ and $M$ the mass of the electron and nucleus, and $L_p^q(t)$ the associated Laguerre polynomials. These $d$-orbitals can be linked with the eigenstates $\{|0\rangle, |1\rangle,|\bar{1}\rangle \}$ of the $L_z$
component of an effective $L=1$ angular momentum with quantum numbers $\{0,1,-1\}$ by the following relations
\begin{eqnarray}
&&\langle \mathbf{r}|0\rangle \rightarrow \phi_{xy}(\mathbf{r})\nonumber \\ 
&& \langle \mathbf{r}|1\rangle \rightarrow \frac{1}{\sqrt{2}} [-i \phi_{xz}(\mathbf{r})-\phi_{yz}(\mathbf{r})] \nonumber \\
&& \langle \mathbf{r}|\bar{1}\rangle \rightarrow \frac{1}{\sqrt{2}} [-i \phi_{xz}(\mathbf{r})+\phi_{yz}(\mathbf{r})] \,. 
\end{eqnarray} 
As done in the main text, $(a,b,c)$ will be used to indicate
the $(d_{yz},d_{xz},d_{xy})$ orbitals.

Now, in order to evaluate the consequence of the electrostatic potential, we need to determine the matrix elements in the Bloch state representation within the same layer and in the neighbors layers along the $z$-direction. These terms will provide, in turn, the amplitude of the orbital Rashba coupling $\alpha_{OR}$ and $\lambda$, respectively.
Let us start by calculating the intra-layer interaction 
\begin{eqnarray}
A^{||}_{p,q}=&& c_{\psi} \langle \psi_{{\bf k},p}({\bf r},i_z) | (-E_s z) | \psi_{{\bf k},q}({\bf r},i_z) \rangle \nonumber \\
=&&  c_{\psi} (-E_s) \frac{1}{N} \sum_{\nu,\gamma} \exp[i {\bf k} \cdot \left({\bf R}_{\nu, i_z}-{\bf R}_{\gamma, i_z}\right)] \times \nonumber \\
&& \times \int d^3 {\bf{r}} \phi^{*}_{p}({\bf r}-{\bf R}_{\nu,i_z})\,z\, \phi_{q}({\bf r}-{\bf R}_{\gamma,i_z}) \,
\end{eqnarray}
\noindent with $p$ and $q$ spanning the orbital index, and $c_{\psi}$ the normalization factor of the Bloch state.
Since the functions $\phi_{p}({\bf r}-{\bf R}_{\gamma,i_z})$ are strongly localized around each atomic position one can restrict the summation to leading terms which are those corresponding to the same site, i.e. ${\bf R}_{\nu, i_z}={\bf R}_{\gamma, i_z}$, and to nearest-neighbor sites, i.e. ${\bf R}_{\nu, i_z}={\bf R}_{\gamma, i_z}\pm {\bf a}_{x,y}$, with ${\bf a}_{x,y}$ being the connecting vectors of nearest-neighbor atoms in the $x-y$ plane. The term for ${\bf R}_{\nu, i_z}={\bf R}_{\gamma, i_z}$ is zero due to the odd-parity symmetry of the atomic functions. 
Then, assuming that the distance between two in-plane nearest-neighbor atoms is $R_{||}$, the amplitude $A^{||}$ can be expressed in a matrix form as 
\begin{eqnarray}
{\hat{A}}^{||}=c_{\psi} (-E_s)\,R_{||}\,I_{||}(R_{||};Z,n) \left[\sin(k_x R_{||}) L_{y} -\sin(k_y R_{||}) L_x \right]
\label{Ainplane}
\end{eqnarray}
\noindent with $I_{||}(R_{||};Z,n)$ being a function of the relative atomic distance $R_{||}$, the atomic number $Z$ and the principal quantum number of the Wannier functions $n$, respectively.
Hence, comparing $A^{||}$ with the term of the Hamiltonian associated with the orbital Rashba coupling, we conclude that the strength of the orbital Rashba coupling $\alpha_{OR}$ is expressed as
\begin{eqnarray}
\alpha_{OR}=(-E_s)\,R_{||}\,I_{||}(R;Z,n) c_{\psi}
\end{eqnarray}
and it is proportional to the intensity of the applied electric field $E_s$ and to the amplitude $I_{||}(R_{||};Z,n)$. The form of ${\hat{A}}^{||}$ in Eq. \ref{Ainplane} is due to the structure of the expectation values of the electrostatic potential between neighbors Wannier functions. 
If we consider schematically the atomic positions $P_A=[0,0,-\frac{R_{\perp}}{2}]$, $P_B=[0,0,\frac{R_{\perp}}{2}]$, $P_C=[R_{||},0,-\frac{R_{\perp}}{2}]$, $P_D=[0,R_{||},-\frac{R_{\perp}}{2}]$, for a cubic geometry in Fig. \ref{fig:schema}(a), we have that
\begin{eqnarray}
\langle\phi_{A,m}| E_s z| \phi_{C,m}\rangle &=&0 \quad {\text{for}}\; m=a,b,c \\
\langle\phi_{A,a}| E_s z| \phi_{C,b} \rangle &=& \langle\phi_{A,b}| E_s z| \phi_{C,c}\rangle =0 \\
\langle\phi_{A,a}| E_s z| \phi_{C,c}\rangle &=& -E_s R_{||} I_{||}(R_{||};Z,n)\\
\langle\phi_{A,c}| E_s z| \phi_{C,a}\rangle &=& +E_s R_{||} I_{||}(R_{||};Z,n) \,.
\end{eqnarray}
\\
The same expressions are obtained along the $y$ directions for the orbitals $b$ and $c$. 
In a similar way, one can proceed for the matrix elements of the electrostatic potential between Bloch states in adjacent layers expressed as
\begin{eqnarray}
A^{\perp}_{p,q}=&&c_{\psi} \langle \psi_{{\bf k},p}({\bf r},i_z) | (-E_s z) | \psi_{{\bf k},q}({\bf r},i_z\pm 1)\rangle \,.
\label{Aoutplane}
\end{eqnarray}
As for the in-plane amplitude, one can expand the summation over all the Bravais lattice. However, in this case there are contributions which are non-vanishing for ${\bf R}_{\nu, i_z}={\bf R}_{\gamma, i_z\pm 1}$ and, thus, we focus on these contributions
\begin{eqnarray}
A^{\perp}_{p,q}=c_{\psi} (-E_s)
\int d^3 {\bf{r}} \phi^{*}_{p}({\bf r}-{\bf R}_{\nu,i_z})\,z\, \phi_{q}({\bf r}-{\bf R}_{\nu,i_z\pm 1}) \,.
\end{eqnarray}
To proceed further we notice that the amplitude $A^{\perp}_{p,q}$ is in general complex because the electric field induces a time dependent vector potential along the $z$-direction that affects the relative phase of the Bloch functions in neighbor layers. This implies that one cannot fix the gauge in a way that the Bloch states in adjacent layers at the surface, e.g. $\psi_{{\bf k},p}({\bf r},i_z=1)$ and $\psi_{{\bf k},p}({\bf r},i_z=2)$, have the same phase. This is an overall phase factor that does not influence the amplitude of the term $A^{\perp}_{p,q}$. Below, we proceed by considering the contribution which leads to a coupling between the electric field and the orbital polarization. 
The form of ${{A}}^{\perp}$ is due to the strucure of the matrix elements of the electrostatic potential between Wannier functions in neighbor layers along the $z-$direction.
Hence, one has to evaluate the following integrals 
\begin{eqnarray}
\int d^3 {\bf{r}} \phi^{*}_{p}({\bf r}-{\bf R}_{\nu,i_z})\,z\, \phi_{q}({\bf r}-{\bf R}_{\nu,i_z\pm 1}) \,.
\end{eqnarray} 
\noindent for nearest neighbor atoms along the $z$-direction as schematically shown in Fig. \ref{fig:schema}.

For the inter-layer term, it turns out that the electric field can induce an orbital polarization on nearest neighbors atoms only if one allows for displacements/distortions of the atoms in the plane with respect to the high-symmetry positions. This physical scenario is  sketched in Fig. \ref{fig:schema}(b,c).
The analysis is performed by considering the following positions for the atoms $A^{'}$ and $B^{'}$ in the plane, $P_{A^{'}}=[-\frac{d_{||}}{2},0,-\frac{R_{\perp}}{2}]$, $P_{A^{'}}=[\frac{d_{||}}{2},0,\frac{R_{\perp}}{2}]$. 
As for the intra-plane case, we have that the relevant non-vanishing integrals are those related to the $L_x$ and $L_y$ components of the angular momentum, namely we have the $L_y$ component that is active for an atomic displacement along the $x$-direction. Within a first order expansion in $d_{||}/R_{\perp}$ one obtains
\begin{eqnarray}
\langle \phi_{A',a} | E_s z |\phi_{B',a} \rangle &=& \langle \phi_{A',b} | E_s z |\phi_{B',b} \rangle = \langle \phi_{A',c} | E_s z |\phi_{B',c} \rangle = 0 \nonumber \\
\langle \phi_{A',a} | E_s z |\phi_{B',b} \rangle &=& 0 \nonumber \\ 
\langle \phi_{A',c} | E_s z |\phi_{B',a} \rangle &=& - \langle \phi_{A',a} | E_s z |\phi_{B',c}\rangle = E_s d_{||} I_{\perp}(R_{\perp};Z,n) \nonumber \\
\langle \psi_{A',b} | E_s z |\psi_{B',c} \rangle &=& 0 \,.
\end{eqnarray}
\\
A similar analysis for a distortive mode along the $y$-direction would give a non-vanishing amplitude only for the wave functions $\phi_{c}$ and $\phi_{b}$. Assuming that the atomic distorsions along the $x$- and $y$-directions have the same amplitude (Fig. \ref{fig:schema}(c)), the resulting expression for the matrix $\hat{A}^{\perp}$ is 
\begin{eqnarray}
\hat{A}^{\perp}=E_s d_{||} I_{\perp}(R_{\perp};Z,n) c_{\psi} (L_x+ L_y) \,.
\end{eqnarray}
\noindent Hence, comparing the structure of $\hat{A}^{\perp}$ with the inter-layer asymmetric interaction introduced in the Hamiltonian, we have that 
\begin{eqnarray}
\lambda=E_s d_{||} I_{\perp}(R_{\perp};Z,n) c_{\psi} \,.
\end{eqnarray}
There are various observations that can be made from the achieved result. Firstly, the inter-layer $\lambda$ coupling is proportional to the applied electric field. Moreover, the electric field penetrating in the skin of the metallic film can couple to the electronic structure by enhancing the orbital polarization through an induced in-plane strain modes. The relative sign of the coefficient in front of the angular momentum operators is not relevant and can be absorbed in the form of the wave function. Moreover, having derived the microscopic expression for $\lambda$ and $\alpha_{OR}$ one can obvserve that their ratio is given by
\begin{eqnarray}
r_{E}=\frac{\alpha_{OR}}{\lambda}=\frac{R_{||}}{d_{||}}\, \frac{I_{||}(R_{||};Z,n)}{I_{\perp}(R_{\perp};Z,n)}
\end{eqnarray}
For a cubic geometry (i.e. $R_{\perp}=R_{||}=R$) and for $n=3$ one can demonstrate that $I_{\perp}(R;Z,n)=I_{||}(R;Z,n)$. However, in thin films $R_{||}$ is typically larger than $R_{\perp}$ due to the vertical confinement and thus the coefficient $I_{\perp}(R;Z,n)$ can be larger than $I_{||}(R;Z,n)$ due to the exponential dependence on the inter-atomic distance.
Furthermore, since the electric field can induce surface strains of the order of 2$\%$ with applied electric field of about $0.3$V/$\AA$\,,\cite{Ben2014} and assuming the differences in the atomic distances, one can estimate $r_{E}$ to have a magnitude varying from about 2 to 20. A detailed quantitative assessment in term of the atomic number and of the inter-atomic distances is beyond the scope of the present manuscript.

\section{Monolayer superconductivity with orbital Rashba coupling}

\begin{figure}[htb]
\begin{center}
\includegraphics[width=0.65\columnwidth]{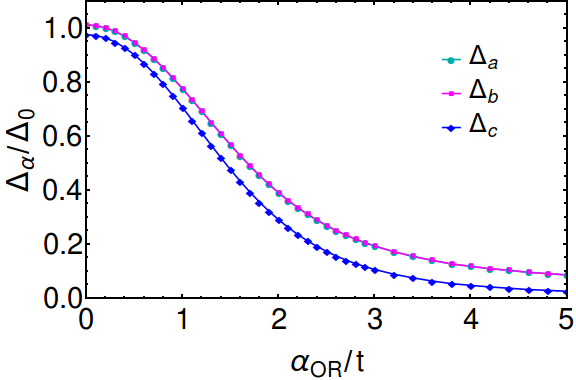} \\
\end{center}
\protect\caption{
Plot of the self-consistent superconducting order parameter $\Delta_\alpha$ ($\alpha=a,b,c$) as function of $\alpha_{OR}/t$ for $g=2.0t$ in a monolayer system.  $\Delta_0$ is the pairing amplitude in absence of OR effect  (i.e for  $\alpha_{OR}=0$) for the 
orbitally isotropic case $(\Delta_a=\Delta_b=\Delta_c)$. 
} 
\label{figSM}
\end{figure}

\begin{figure}[bth]
\includegraphics[height=4.4cm]{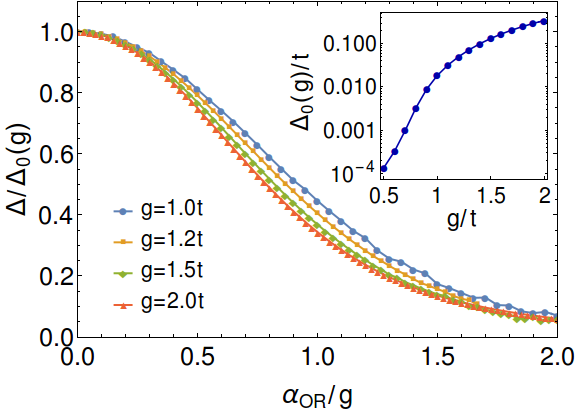}
\protect\caption{Behavior of the superconducting order parameter (OP) $\Delta$ in a monolayer  as a function of the orbital Rashba interaction parameter  ($\alpha_{OR}$), for different values of the pairing strenght $g$. The OP is here assumed orbitally-uniform, i.e. with $\Delta_a=\Delta_b=\Delta_c$. We observe that the smaller the value of $g$ the stronger the effect of suppression and reduction of the OP due to the orbital Rashba effect. Inset: Behavior of the SC OP in a monolayer in absence of OR effects ($\Delta_0$) as a function of $g$. This value is used as a scale for $\Delta$.
} 
\label{fig:gOP}
\end{figure}

For a monolayer configuration, the presence of the orbital Rashba (OR) coupling tends to reduce the strength of the superconductivity by inducing a suppression of the order parameter (OP). This behavior is explicitly demonstrated in Fig.~\ref{figSM}, where the superconducting OP amplitude for each band, self-consistently determined, exhibits a monotonous decrease as a function of  $\alpha_{OR}$. 
We have also considered the self-consistent value of $\Delta_{\alpha}$ while changing $\alpha_{OR}$ assuming different values of the pairing coupling $g$. 
We find that the amplitudes are in general scaled by means of the pairing coupling $g$ and thus in the following and in the main text we have performed the calculation assuming $g=2.0\,t$. The scaling behavior is reported in the Fig. \ref{fig:gOP}.

The behavior of the order parameter is almost collapsing on the same curve as a function of $\alpha_{OR}/g$. This result implies that the amplitude of the superconducting gap on the surface substantially depends on the ratio between the orbital Rashba coupling and the pairing interaction strength.  
\\

\section{Competing phases and character of the phase transitions}

In this section, we study the free energy of the examined model Hamiltonian by considering the order parameter $\Delta_\alpha(i_z)$ as uniform through the layers and isotropic in the orbital channels. The analysis is done by introducing the variable $\Delta$ which is the common amplitude of the OPs in the various orbital channels, i.e. $|\Delta_a|=|\Delta_b|=|\Delta_c|=\Delta$, while the sign is added when evaluating the $\pi$-SC configuration
(i.e. $\Delta_c=-\Delta_{a,b}=-\Delta$).

\begin{figure}[th]
\begin{center}
\includegraphics[width=0.45\columnwidth]{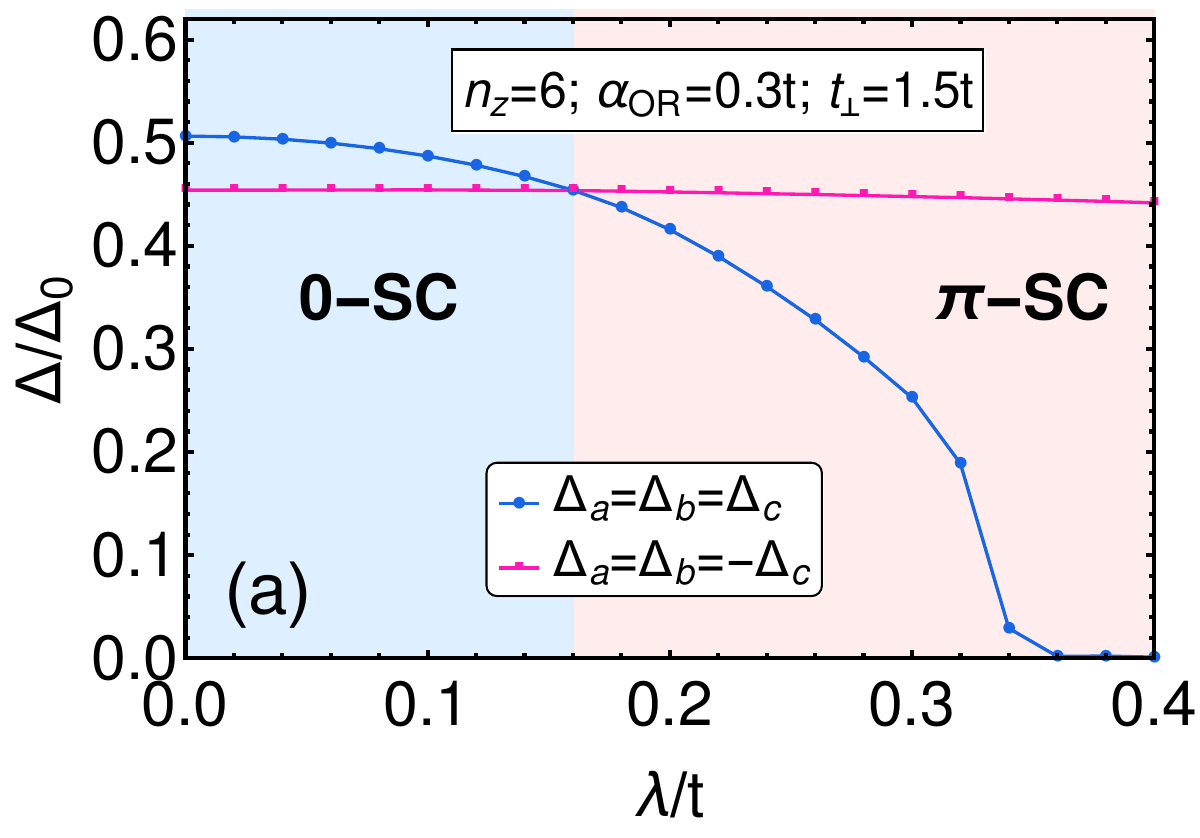}
\includegraphics[width=0.45\columnwidth]{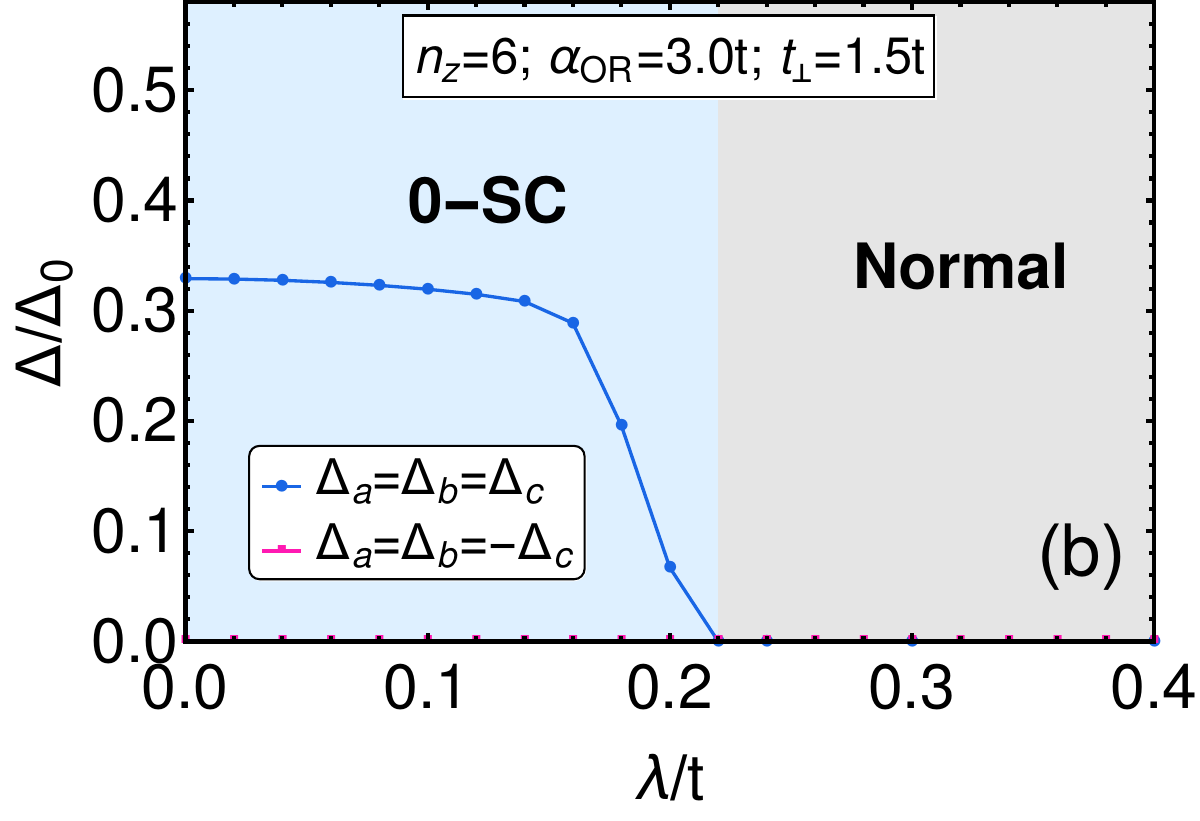} \\
\vspace{0.3cm}
\includegraphics[width=0.45\columnwidth]{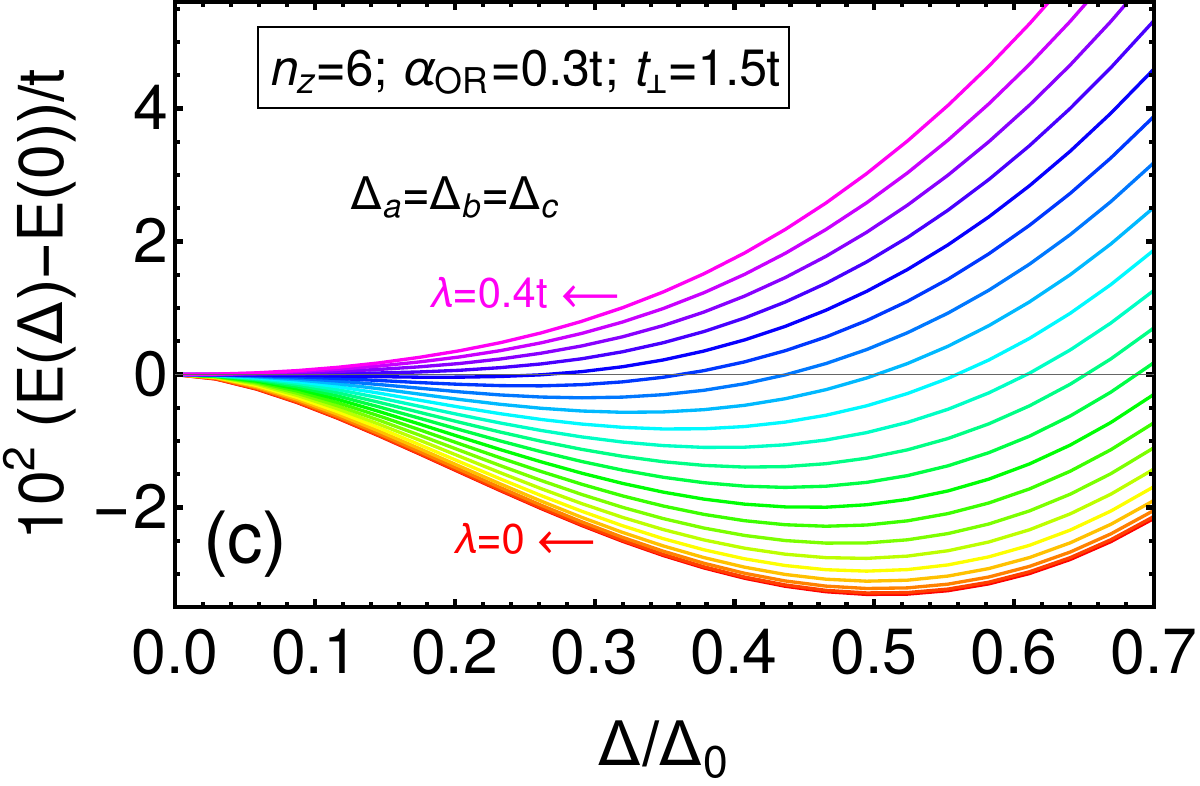}
\includegraphics[width=0.45\columnwidth]{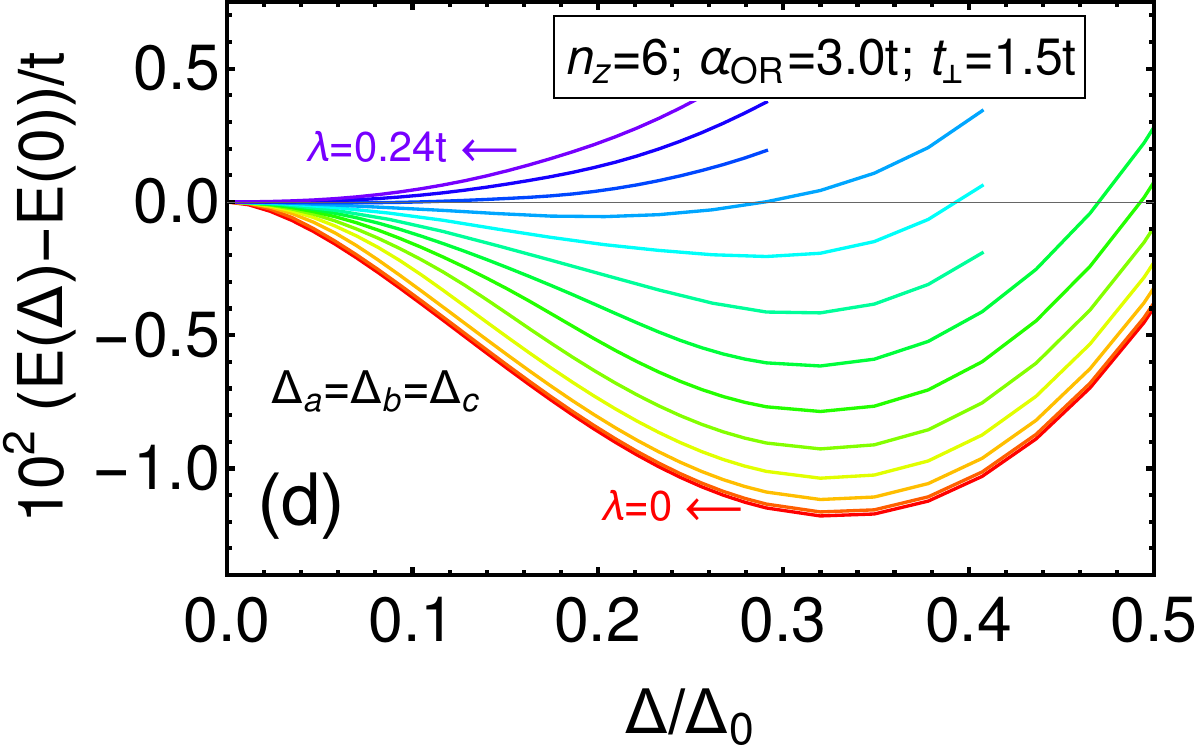} 
\\
\vspace{0.3cm}
\includegraphics[width=0.45\columnwidth]{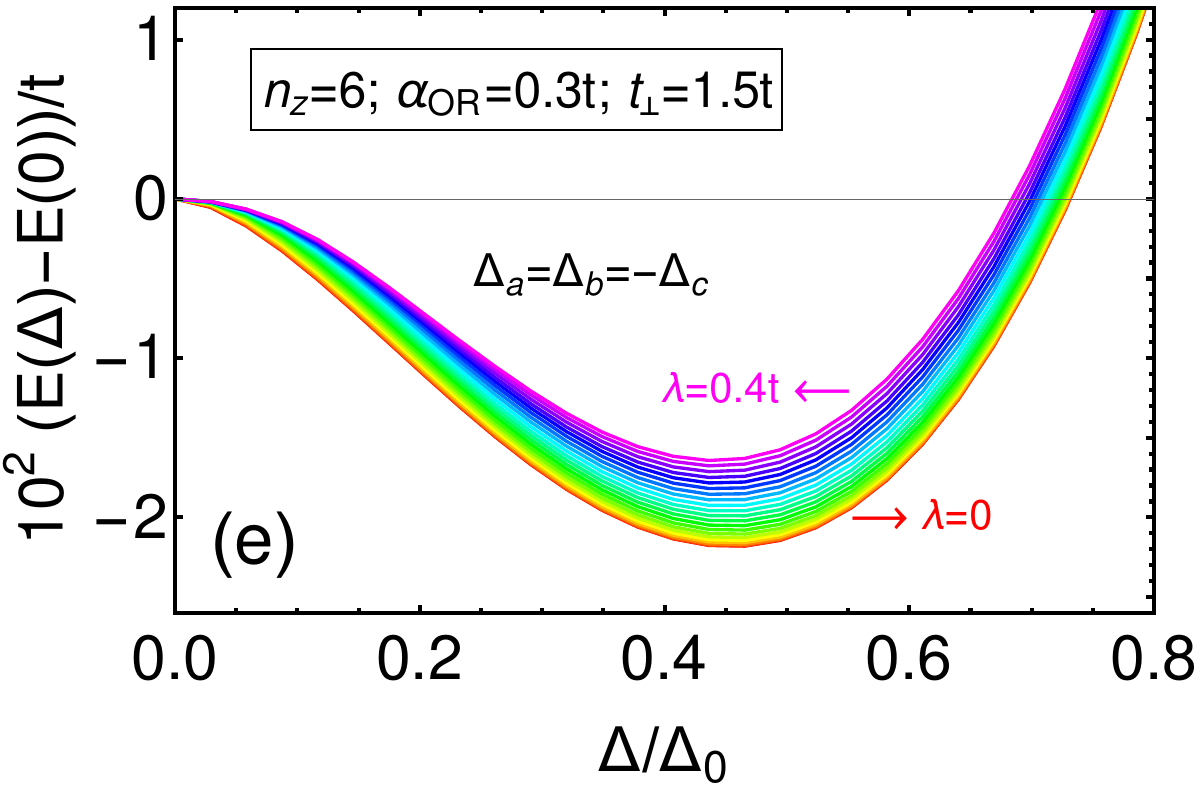}
\includegraphics[width=0.45\columnwidth]{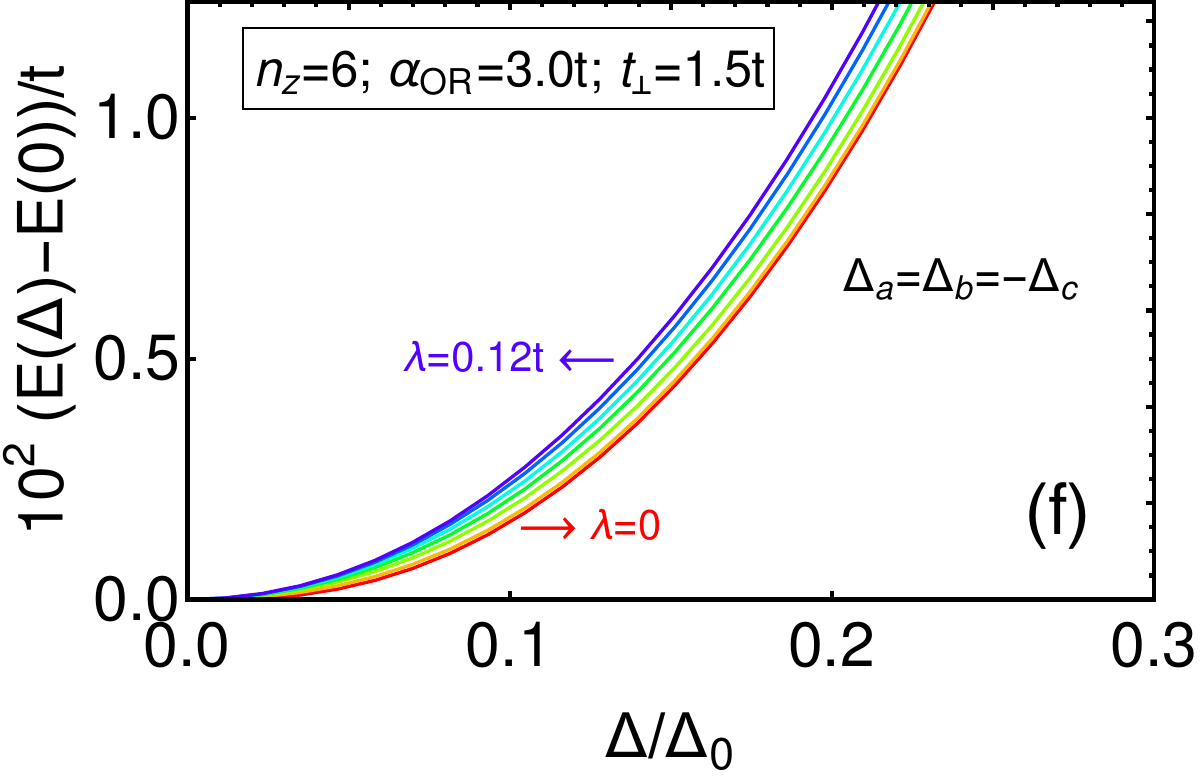} 
\end{center}
\protect\caption{(a)-(b) Plot of the value of $\Delta$ that minimizes the free energy varying $\lambda$ (with $|\Delta_a|=|\Delta_b|=|\Delta_c|=\Delta$), for weak (panel (a)) and strong (panel (b)) orbital Rashba strength. The blue line corresponds to the analysis in which $\Delta_\alpha$ ($\alpha=a,b,c$) have all the same sign, the magenta line instead refers to a configuration with $\Delta_c$ having an opposite sign with respect to $\Delta_{a,b}$. We observe that in (a) we have a transition (around $\lambda\simeq 0.16t$ to the $\pi$-SC state). (c)-(f) Plots of the free energy as a function of $\Delta$ for several values of $\lambda$ in different cases, namely: (c) for weak $\alpha_{OR}$ and $\Delta_\alpha >0$ with  $\lambda/t \in[0,0.4]$; (d) for strong $\alpha_{OR}$ and $\Delta_\alpha >0$ and $\lambda/t \in [0,0.24]$; (e) for weak $\alpha_{OR}$ and $\Delta_c<0$ and $\lambda/t \in [0,0.4]$; and finally in (f) for strong  $\alpha_{OR}$ and $\Delta_c<0$ and $\lambda/t \in [0,0.12]$. For the four free energy plots the different lines refer to inequivalent values of $\lambda$, for the reported ranges, starting from $\lambda=0$ (red bottom line), up to the highest value of $\lambda$ analyzed, following the rainbow colors with step $0.02t$. } 
\label{figSM2}
\end{figure}

 Representative cases of weak and strong OR effect, namely with $\alpha_{OR}=0.3t$ and $\alpha_{OR}=3.0t$ respectively, are reported in Figs.~\ref{figSM2}. We have considered a system with $n_z=6$ layers and with interlayer hopping $t_\perp=1.5t$. Similarly to the full self-consistent analysis (see Fig.~2 in the main text), we find that  the increase of $\lambda$ drives a transition between 0-SC and $\pi$-SC states for weak $\alpha_{OR}$ (see Fig.~\ref{figSM2}(a)), and a transition from SC to normal state for strong $\alpha_{OR}$ (see Fig.~\ref{figSM2}(b))). Indeed, for $\alpha_{OR}=0.3t$ 
comparing the panels (c) and (e) of Fig.~\ref{figSM2}, we see that for $\lambda>0.16t$ the case with $\Delta_c<0$ has an energetically more favorable solution. This transition if of first order, since we have a discontinouity in the first derivative of the free energy.

For larger values of $\alpha_{OR}$, the free energy of the case with $\Delta_c<0$ has a minimum only for $\Delta=0$ (i.e. normal state solution) as can be easily deduced from Fig.~\ref{figSM2}(f). Hence, for $\alpha_{OR}=3.0t$ the system never reaches the $\pi$-SC phase and we observe a continuous transition from SC to normal state by following the free energy minima, as shown in Fig.~\ref{figSM2}(b) and (d), for $\lambda \simeq 0.21t$.
The values of the transition points are slighlty different from those reported in the phase diagram (Fig.~2(a) of the main paper), since in the present analysis we are assuming an uniform and isotropic superconducting OP.

\begin{figure}[th]
\includegraphics[width=0.9\columnwidth]{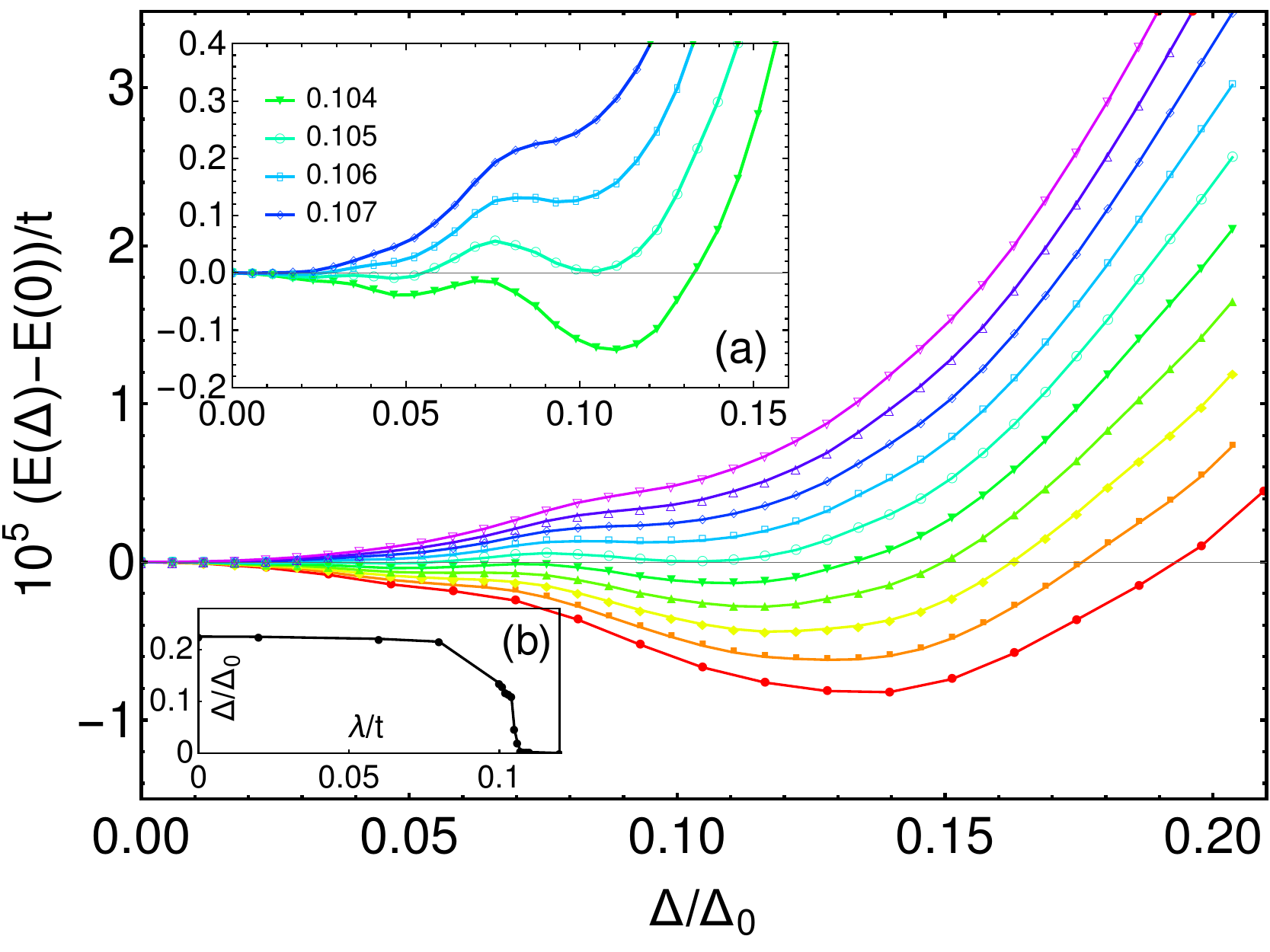}
\protect\caption{Plot of the energy difference ($E(\Delta)- E(0)$) as a function of the order parameter $\Delta$ (assuming that $|\Delta_a|=|\Delta_b|=|\Delta_c|=\Delta$) for several values of $\lambda$ close to the critical point for strong OR coupling ($\alpha_{OR}=3.0t$). The system has $n_z=6$ layers and $t_\perp=1.5t$. The analysis has been made by assuming that the superconducting order parameter is zero in the outer layers and uniform in the inner ones.  In the inset (a) there is a zoom for four values of $\lambda$, underlying the presence of multiple minima and hence the occurrence of a first order phase transition. (b) Behavior of  the order parameter as a function of the orbital mixing term $\lambda$. After the small discontinuity the second minimum goes smoothly to zero.}
\label{figSM3}
\end{figure}

Indeed, for strong OR effect, the uniform profile of the OP within the whole superconductor is not a good assumption since the values of $\Delta_\alpha(i_z)$ in the outer layers are strongly suppressed, compared to those in the inner layers, as can be seen in Fig.~\ref{figTP}(c)-(d). 
Hence, for $\alpha_{OR}=3.0t$ we have also performed an analysis in which we assume that the order parameter is zero in the outer layers and uniform in the remaining ones. Results are reported in Fig.~\ref{figSM3}, where we observe the presence of multiple minima and the increase of $\lambda$ drives a weak first order transition before the continuous second order SC-normal transition is achieved. 

\section{Layer dependent orbital polarization}

\begin{figure*}[th]
\includegraphics[width=0.9\textwidth]{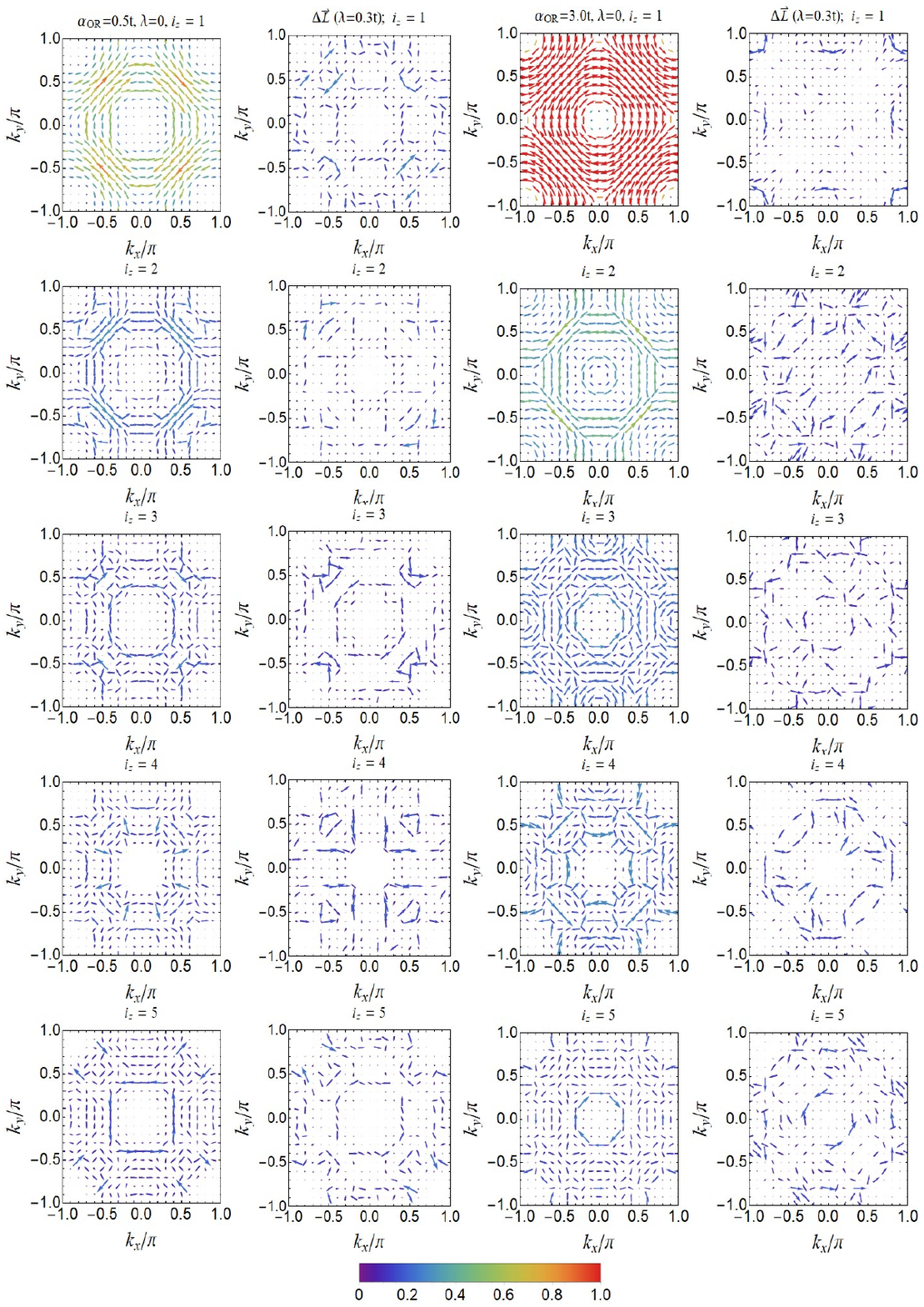}
\protect\caption{Vector plots of $\bm L(k_x,k_y)$ per layer in the Brillouin zone, for a superconducting system with 10 layers, in the case of weak ($\alpha_{OR}=0.5t$) and strong ($\alpha_{OR}=3.0t$) orbital Rashba effect. $L_z=0$ everywhere. In the first and third column we show the vector plots for $\lambda=0$, while in the second and fourth we show the difference $\Delta\bm L$ between the vectors $\bm L$ for $\lambda=0.3t$ and $\lambda=0$. 
In each row we present the behavior in each layer labeled by $i_z$. Since the system is symmetric, layers from 6 to 10 are not shown.  The arrows are colored according to the magnitude of the vector field (see the legend at the bottom), with intensities that are scaled to unity. In each panel, the magnitude of $\bm L$ is also represented by the dimension of the arrows.}
\label{figOP2}
\end{figure*}

Finally, we present the layer dependent orbital polarization for a superconducting heterostructure with $n_z=10$. The analysis is performed by considering firstly the role of the OR coupling at the surface and how the obtained orbital polarization in the Brillouin zone is also transferred inside the inner layers (first and third column of Fig. \ref{figOP2}). Starting from the case at $\lambda=0$, one can observe a chiral texture of the orbital components with windings around the high symmetry points of the Brillouin zone (BZ). In particular the winding around the $\Gamma$ point at $(k_x,k_y)=(0,0)$ is opposite to that occuring around the $M$ point at $(\pi,\pi)$ with a domain wall in between the $(0,\pm\pi)$ and $(\pm \pi,0)$ points. We observe that moving from the surface to the inner layers, the domain walls proliferate and there are extra structures emerging along the diagonal of the BZ with opposite orbital chirality. We notice that the presence of the orbital Rashba coupling at the surface is sufficient to induce a non-trivial orbital polarization into the inner layers of the superconductor (see first and third columns of Fig. \ref{figOP2}).  

The effect of $\lambda$ is then investigated by evaluating the difference in the orbital texture with respect to the configurations with $\lambda=0$ by keeping the samle amplitude of $\alpha_{OR}$.  
As one can see in the second and fourth column of Fig. \ref{figOP2}, the effect of $\lambda$ is to amplify the formation of pockets of orbital textures with inequivalent or opposite orientation of the orbital polarization thus indicating an orbital connectivity which is less regular if compared to the case without $\lambda$. Such structure of the orbital texture in the reciprocal space contributes to reduce the superconducting pairing which is maximally favored for electron pairs without any orbital polarization.


%


\end{document}